\documentclass[12pt,preprint]{aastex}

\newcommand{\einstein}{{\sl Einstein\/}}
\newcommand{\exosat}{{\sl EXOSAT\/}}
\newcommand{\ginga}{{\sl Ginga\/}}
\newcommand{\rosat}{{\sl ROSAT\/}}
\newcommand{\asca}{{\sl ASCA\/}}
\newcommand{\sax}{{\sl BeppoSAX\/}}
\newcommand{\xte}{{\sl RXTE\/}}
\newcommand{\cv}{V1432~Aql}
\newcommand{\agn}{NGC~6814}
\newcommand{\pspin}{P$_{\rm spin}$}
\newcommand{\porb}{P$_{\rm orb}$}
\newcommand{\pbeat}{P$_{\rm beat}$}
\newcommand{\cps}{cts\,s$^{-1}$}

\shorttitle{X-ray Variability of V1432 Aql}
\shortauthors{Mukai et al.}

\begin{document}

\title{X-ray Variability of the Magnetic Cataclysmic Variable V1432 Aql \\
       and the Seyfert Galaxy NGC 6814}

\author{K. Mukai\altaffilmark{1,2}}
\affil{Code 662, NASA/Goddard Space Flight Center, Greenbelt, MD 20771, USA}
\email{mukai@milkyway.gsfc.nasa.gov}

\author{C. Hellier}
\affil{Department of Physics, Keele University, Keele, Staffordshire
	ST5 5BG, United Kingdom}

\author{G. Madejski}
\affil{Stanford Linear Accelerator Center, Glast Group, 2575 Sand Hill Road,
	MS 43A, Menlo Park, CA 94025, USA}

\author{J. Patterson}
\affil{Department of Astronomy, Columbia University, 538 W. 120th Street,
	New York, NY 10027, USA}

\and

\author{D.R. Skillman}
\affil{Center for Backyard Astrophysics (East), 9517 Washington Avenue,
	Laurel, MD 20723, USA}

%\email{aastex-help@aas.org}

\altaffiltext{1}{Also Universities Space Research Association}
\altaffiltext{2}{Also Columbia Astrophysics Laboratory, Columbia University,
	550 West 120th Street, New York, New York 10027, USA}

\begin{abstract}

V1432~Aquilae (=RX~J1940.2$-$1025) is the X-ray bright, eclipsing
magnetic cataclysmic variable $\sim$37$'$ away from the Seyfert
galaxy, \agn.  Due to a 0.3\% difference between the orbital
(12116.3 s) and the spin (12150 s) periods, the accretion geometry
changes over the $\sim$50 day beat period.  Here we report the
results of an \xte\ campaign to observe the eclipse 25 times,
as well as of archival observations with \asca\ and \sax.
Having confirmed that the eclipse is indeed caused by the secondary,
we use the eclipse timings and profiles to map the accretion geometry
as a function of the beat phase.  We find that the accretion region is
compact, and that it moves relative to the center of white dwarf on
the beat period.  The amplitude of this movement suggest a low-mass
white dwarf, in contrast to the high mass previously estimated from
its X-ray spectrum.  The size of the X-ray emission region appears
to be larger than in other eclipsing magnetic CVs.
We also report on the \xte\ data as well as the
long-term behavior of \agn, indicating flux variability by a factor
of at least 10 on time scales of years.
\end{abstract}


\section{Introduction}

The Seyfert galaxy \agn\ became famous with the apparent discovery,
made with the non-imaging \exosat\ Medium-energy Experiment (ME),
that its X-ray flux was modulated with a strict $\sim$12,150 s period
\citep{MB1989}.  However, this fame turned into notoriety when
it was discovered, through imaging X-ray observations with \rosat\ Position
Sensitive Proportional Counter (PSPC), that this periodicity actually
belonged to another X-ray source in the field of view, RX~J1940.1$-$1025,
about 37$'$ away from \agn\  \citep{Mea1993,Sea1994}.  The latter authors
identified this new X-ray source with a V$\sim$16 mag object, subsequently
designated V1432~Aquilae, and proposed that it was a cataclysmic variable
(CV) system belonging to the polar sub-class (also known as AM Her-type
systems).

A CV is a semi-detached binary in which the white dwarf primary
accretes mass from a late-type (usually an M dwarf) secondary.
In a polar (see \citealt{C1990} for a review), the primary is strongly
($\geq$10 MGauss) magnetic, which prevents the formation of an accretion
disk.  The accretion flow is channeled by the magnetic field towards the
magnetic pole region(s), where a strong standing shock forms.  The post-shock
plasma cools by emitting optical/infrared cyclotron emission and optically
thin thermal X-ray emission with a typical temperature of kT$\sim$10 keV
(``the hard component'').  The white dwarf surface is heated from above
by these radiation, and probably directly by dense blobs whose shocks are
buried within the white dwarf atmosphere, and emit optically thick,
blackbody-like emission with a typical temperature of kT$\sim$30 eV
(``the soft component'').  In most polars, the soft X-ray component
dominates over the hard, although in some the two are roughly comparable.

In polars, the strong magnetic field of the primary usually synchronizes
the white dwarf spin to the orbital period.  In the other subclass
of magnetic CVs, intermediate polars (IPs) or DQ Her-type systems
(see \citealt{P1994} for a review), the spin period of the white dwarf
(\pspin) is significantly shorter than the orbital period (\porb), many
of them with \pspin\ $\sim$ 0.1 \porb.  In IPs, a weaker magnetic field
or a greater orbital orbital separation is thought to prevent synchronism,
and that their current \pspin\ is close to their evolutionary equilibrium.

\cv\ is among the minority of polars with a strong hard X-ray component.
Moreover, two almost identical periods have been found in this system
\citep{Pea1995,Fea1996}.  The orbital period is 12116.3 s, while the
second period is $\sim$0.3\% longer at 12150 s, with a resulting beat
period (\pbeat$^{-1}$ = \porb$^{-1}$ $-$ \pspin$^{-1}$) of $\sim$50 days.
This makes \cv\ an asynchronous polar, in which \pspin\ and \porb\ differ
by $\sim$1\%, probably in a temporary departure from synchronism.
In the first known example, V1500~Cyg (=Nova Cygni 1975), the nova explosion
is the almost certain cause of the asychronism.  In an asynchronous polar,
a strong ($\gg$accretion torque) MHD torque should operate to synchronize
the white dwarf spin on a short ($\ll$evolutionary) time scale, and indeed
this is observed in V1500~Cyg (synchronization timescale $\sim$150 years;
\citealt{Sea1995}).  In addition to V1500~Cyg, BY~Cam and CD~Ind are also
asynchronous with \pspin\ being $\sim$ 1\% shorter than \porb.
On the other hand, \cv\ appears to be asynchronous in the other
direction.  \cite{M1998} questioned if this was possible, and also
pointed out that such a system might be expected to synchronise much
more quickly than V1500~Cyg like systems: When \porb\ $>$ \pspin,
the synchronization must compete and overcome the accretion torque,
whereas when \porb\ $<$ \pspin, they cooperate.  Although \cite{SS2001}
showed that it was possible for a nova explosion to slow down the white
dwarf spin, thus producing a \cv-like system, the synchronization timescale
of \cv\ that \cite{GS1997} inferred is 100 yrs,
comparable to that of V1500~Cyg.

\cite{M1998} proposed an alternative model in which \pspin\ is about
4040 s.  A complex interaction of spin and orbital periodicity, and
less than perfect observational sampling could explain the apparent
presence of a 12150 s period.  However, we now consider this model
unlikely, given the confirmation of rapid timescale for synchronization
(\S 4.1).  We will therefore refer to the 12150 s period as the spin period
in the rest of this paper.

There is a second controversy surrounding \cv.  The fiducial marker
that defines the 12116.3 s orbital period is an eclipse-like event
seen both in the optical and X-ray light curves.  However, \cite{Wea1995}
argued this to be a dip caused by the accretion stream, as seen in several
well-observed polars \citep{KW1985}.  They based this conclusion partly
on the light curve which is unusual for an eclipsing polar, but mainly
on the residual X-ray flux observed with non-imaging \ginga\ Large
Area Counter (LAC) detector.  The observed spectrum became harder during
the eclipse, which suggested a photoelectric absorption event by the
accretion stream.  However, the presence and the regularity of the
same feature at all energies strongly argues that this is a true
eclipse by the secondary star \citep{Pea1995}.  In this paper, we
show that later X-ray observations are consistent with zero residual
flux, thus strongly favoring the eclipse interpretation.  We then
go on to use the eclipse as a tool to explore the accretion geometry
as a function of the beat phase.

Although not the origin of the $\sim 12000$ s period, the Seyfert galaxy
\agn\ is by no means an uninteresting object.  In the optical, it shows
broad emission lines of only modest width ($\sigma \sim 3000$ km\,s$^{-1}$);
both the continuum \citep{D1988} and the emission lines \citep{SM1990} are
highly variable, varying between Seyfert 1 or Seyfert 1.8 classifications.
The \rosat\ observations show that the X-ray flux of \agn\ is also variable,
decreasing by a factor of at least 5 between October 1992 and October 1993,
and by a factor of 2 in $\sim$ 6 hours \citep{Kea1997}.  Given the angular
separation of $\sim 37'$ between \cv\ and \agn, X-ray data with
non-imaging, collimated instruments are often entangled; the strong
variability of \agn\ then leads to require particular care to be taken
when observing \cv\ with such an instrument.  Therefore, although \cv\ is
the primary focus of this work, we also present results on the X-ray
behaviors of \agn, both as a necessary precondition for the study of
the CV and as a worthy goal in itself.

We describe the X-ray observations in \S 2.  We present the results
and implication on \agn\ in \S 3.  The results on \cv\ (including a
short summary of continued optical photometry) are presented in \S4,
and discussed in \S 5.  A summary of conclusions is given in \S 6.

\section{X-ray Observations}

In addition to our own \xte\ campaign, we have analyzed archival
observations with two X-ray observatories with imaging telescopes.
All X-ray observations analyzed here were obtained during 1997 and 1998.

\subsection{\sax\ Observations}

\cv\ was observed on 8 occasions between 1997 April 2 and May 13
with \sax\ \citep{Bea1997a}.  Here we will focus on the data from
Medium Energy Concentrator Spectrometer (MECS) instruments \citep{Bea1997b}.
\sax\ carried 3 identical units of MECS, but one had malfunctioned between
the 7th and 8th observations of \cv, so the last observation was made with
only two operational MECS.  Each visit resulted in between 8 and 13 ksec
of effective exposure with the MECS.  Details of the 8 observations
are listed in Table\,\ref{obslogsa}.

Although the Low Energy Concentrator Spectrometer (LECS) can in principle
provide information regarding the behavior of the source below 2 keV, the
data quality is lower than the MECS because there is only one unit of LECS
and it is operated only during orbital night.  Moreover, the latter means
that LECS data sample a subset of the phase covered by MECS data.  Given
the potential complications from cycle-to-cycle variability, we have opted
not to include LECS data in our analysis.  Similarly, we have not included
data from the two non-imaging narrow-field instruments on-board \sax,
the High Pressure Gas Scintillation Proportional Counter and the Phoswich
Detection System.

We have extracted source counts from the MECS data using a circular
extraction region with a radius of 3.6 arcmin, and background counts
from an annular extraction region with outer and inner radii of 16.9
and 7.2 arcmin, respectively, assumed that the background was flat,
and therefore scaled the latter by the ratio of areas of the extraction
regions to derive the net source count rates.  We have extracted light
curves from PI channels 31--100 ($\sim$1.5--4.5 keV) and channels
101--170 ($\sim$4.5--8.0 keV) with a bin size of 16 s.  Roughly speaking,
these two bands contain equal counts from \cv, and when combined,
result in the highest signal-to-noise.

\subsection{A Serendipitous \asca\ Observation}

\cv\ was observed serendipitously with \asca\ in 1997 October
during the observation of a high redshift quasar, PKS~1937$-$101
(see Table\,\ref{obslogsa} for further details).
Although \asca\ has 4 co-aligned telescopes \citep{Tea1994},
\cv\ was about 20 arcmin off-axis and was not in the field of view
of the two Solid-state Imaging Spectrometers.  As for the two Gas Imaging
Spectrometers (GIS), the source was too close to the on-board radioactive
calibration source in GIS-3 for a useful analysis.  We therefore
restrict ourselves to the GIS-2 data.

We have extracted the source counts from an elliptical extraction
region (to match the overall shape of the off-axis point spread
function) with semi-major and semi-minor axes of 7.5 and 2.9 arcmin,
respectively.  Background was taken from a large area of the detector
with a similar off-axis angle as the source, which was then subtracted
from the source data after scaling with the ratio of the detector areas.
We have used two energy bands, PI channels 60--169 (0.7--2 keV) and
170--847 (2--10 keV) for the \asca\ data.

\subsection{The \xte\ Campaign}

We have performed a series of 25 observations between 1998 May 18
and 1998 Aug 21, covering approximately two beat cycles of \cv\ with
\xte\ \citep{Bea1993}.  Each observation covers a small range of
orbital phase of \cv, centered on the eclipse.  Here we concentrate
on the data obtained with the Proportional Counter Array (PCA)
\citep{Jea1996}.

The angular separation of \cv\ and \agn\ of $\sim$37 arcmin means that
PCA observations pointed at \cv\ have some sensitivity to the X-ray
photons from \agn.  We have therefore chosen to offset our pointing
by $\sim$15$'$ (position ``C'' in Table\,\ref{obsposxte} and in
Figure\,\ref{map}).  We estimate that the PCA collimater response is
about 80\% for \cv\ and about 10\% for \agn\ at position C.  Furthermore,
we wished to monitor the X-ray flux level of \agn\ to assess the residual
contamination, and for its own scientific merits.  Our intention was to
similarly offset the pointing.  However, due to a typographical error on
our part, the actual pointings were performed at position A in
Table\,\ref{obsposxte} and in Figure\,\ref{map} (change the minute figures
of RA to 43 and we would have achieved our original aim; position A'
in Figure\,\ref{map}). This greatly reduced the effectiveness of our campaign
for \agn\ (see, however, \S 3).

Each of the 25 observations was composed of a pointing at position C
preceded by, followed by, or in two cases sandwiched between, a pointing
at position A (see Figure\,\ref{xteagn} for a couple of examples).
Typically 2 ksec of good data were obtained at position C.
The details of the \xte\ observations are given in Table\,\ref{obslogxte}. 
As is often the case, not all of the five Proportional Counter Units (PCUs)
were active in all observations: the active PCUs are also indicated in the
table. The collimator response at off-axis positions are a function of the
roll angle, since the collimator geometry is hexagonal.  We have estimated
the actual efficiency achieved in each C pointing for \cv\ and for \agn,
so that we can infer the on-axis count rates (per second per PCU) from the
observed rates.  We have estimated the background using faint source
background model, using files pca\_bkgd\_faint240\_e03v03.mdl and
pca\_bkgd\_faintl7\_e03v03.mdl as well as the South Atlantic Anomaly
history file.  This model is highly successful in reproducing the non-X-ray
background experienced by the PCA.  However, it only models the average,
high-latitude, cosmic X-ray background, which is not appropriate for this
part of the sky.  Moreover, the model does not include the contamination
by \agn.

We have extracted data from the top layer of the PCA, which gives
the best signal-to-noise ratio.  For the study of \agn, we use
the entire energy band of the PCA (effectively 3--15 keV), while
for the detailed study of the eclipses in \cv, we restrict ourselves
to channels 8--27, or roughly 3--10 keV, which on average improves
the signal-to-noise ratio.  In both cases, we use count rate per PCU
in our plots to account for the variable number of active PCUs,
and moreover convert this to the estimated on-axis rate for the
appropriate target.

\section{X-ray Variability of \agn}

\subsection{The Medium Energy X-ray Light Curves of \agn}

We now use our \xte\ campaign as well as archival data
to evaluate the X-ray fluxes of \agn\ in the 2--10 keV band.
Lower energy X-rays from \agn\ have been securely detected
using imaging instruments and are known to be variable,
starting with the \einstein\ Imaging Proportional Counter (IPC)
observation spanning two days, which showed this source to be variable
on a timescale of 20,000 s \citep{T1980}.  A long-term
light curve obtained with \exosat\ Low-energy Experiment (LE) is presented
in Figure 1 of \cite{MB1989}; and for \rosat\ results, see \cite{Kea1997}.
As \cite{MB1989} noted, the \exosat\ Medium-energy Experiment (ME)
data of ``\agn'' varied in a manner similar to that of the LE data
over the timescale of months (their Figure 1).  Although we now know
that the (non-imaging) ME data were contaminated with X-rays from \cv,
as evidenced by the detection of the 12,000 s signal, the LE--ME correlation
is a strong evidence that a significant fraction of the observed ME counts
did originate from \agn.

In our \xte\ campaign, we attempted to monitor the X-ray variability of
\agn\ by obtaining data at position ``A'' as well as at position ``C.''
Although our typographical error made it less effective than we had hoped for,
we believe we have securely detected \agn\ during two pointings,
on 1998 May 18 and on 1998 June 16.  The light curves on these two dates
at both pointing positions are plotted in Figure\,\ref{xteagn}.
In both cases, the curve is flat at position ``A'' at a level higher
than the adjacent section of light curve taken at ``C.''  We also show,
in Figure\,\ref{xteagnspec}, the spectra on 1998 May 18 taken at these two
pointing positions.  The ``A'' spectra on May 18 and June 16 can both be
fit with a single power law without an obvious Fe K$\alpha$ line, with
a photon index of $\sim$1.9.  The inferred 2--10 keV fluxes were 3.4 and
4.9 $\times 10^{-11}$ ergs\,cm$^{-2}$s$^{-1}$, respectively, although this
assumes all the photons were from the \agn\ (the actual fluxes were probably
somewhat lower, since some contamination by \cv\ is likely).

We have another method to estimate the X-ray flux from \agn.
By interpreting the mid-eclipse residual count rates from ``C'' observations
(see \S 4.3 for details) as off-axis contribution from \agn,
we have constructed its light curve during our \xte\ campaign
(Figure\,\ref{agnmon}), although we cannot rule out other contributors
to the mid-eclipse counts (including residual flux in \cv\ itself).
On the two dates where the ``A'' pointing securely detected \agn,
the count rates measured by this method are somewhat lower, and in fact may
be more realistic.  Other points on this curve lack independent confirmations,
but nevertheless may be considered an estimate of the level of the 2--10 keV
flux variability of \agn\ during the \xte\ campaign.

Since the 2--10 keV flux indicated for \agn\ in 1998 appeared much higher
than expected, given the \rosat\ results \citep{Kea1997}, we have investigated
the long-term flux history of this object further.  The most reliable
determination of its 2--10 keV flux can be obtained from the 1993 May
\asca\ observation: the source was detected at 0.044 \cps\ in the GIS,
had a power law spectrum with a photon index of 1.7 and Galactic absorption
(9.8$\times 10^{20}$ cm$^{-2}$; \citealt{Eea1989})
and a 2--10 keV flux of 1.7 $\times 10^{-12}$ ergs\,cm$^{-2}$s$^{-1}$.
We use this spectral shape to infer 2--10 keV flux from lower-energy
imaging observations (\einstein\ IPC and High Resolution Imager observations
in 1979 April, \exosat\ LE observations in 1983--1985, the \rosat\ all-sky
survey data in 1990, and pointed \rosat\ PSPC observations in 1992 and 1993).
Although extrapolation can lead to inaccuracies, if the spectral shape was
different (e.g., with a soft excess) during any of these observations,
our estimates are probably accurate to within a factor of 2.  We also note
that \agn\ was detected at 0.36 \cps\ in the \rosat\ All-Sky Survey
(http://www.rosat.mpe-garching.mpg.de/survey/rass-bsc/; the observation
dates were estimated using the formula given in \citealt{Vea1997}).
This extends the range of PSPC count rates of \agn\ to a full
order of magnitude from a factor of 5 noted by \cite{Kea1997}.
In Figure\,\ref{agnlong}, we plot the reconstructed
flux history of \agn, including, additionally, the two most reliable (but
also among the brightest) values from our \xte\ campaign, and two additional
points based on mid-eclipse flux during the 1989 and 1990 \ginga\ observations.
This shows a wide range in estimated flux, from $\sim 1.0 \times 10^{-12}$ to
$> 1.0 \times 10^{-11}$ ergs\,cm$^{-2}$s$^{-1}$.  The highest point is
based on \exosat\ LE rate of 0.0225 \cps\ on 1983 September 4, at
an estimated 2--10 keV flux of $3.3 \times 10^{-11}$ ergs\,cm$^{-2}$s$^{-1}$,
which accounts for most of the ME count rates observed on that occasion.
In the 1985 October 16 observation in which the 12,000 s period was
discovered, we estimate that \agn\ at $\sim 1.0 \times 10^{-11}$
ergs\,cm$^{-2}$s$^{-1}$ (2--10 keV), and infer that the observed ME
counts were roughly half from the AGN and half from the CV.  This is
consistent with the fact that ME count rate never dropped below
$\sim$0.8 \cps, even during the eclipse of \cv.

Finally, we estimate that our \xte\ observing strategy was successful
in greatly reducing the \agn\ contributions to the count rates when our
intended target is \cv.  Figure\,\ref{agnmon}, multiplied by the collimator
efficiency for \agn\ for each C poiting ($\sim$0.1), can be interpreted as
the upper limit to the contamination level due to \agn.  The AGN and the
Galactic X-ray background prevent us from setting a strict upper limit to
the flux at mid-eclipse to better than $\sim$0.15 \cps\ per PCU.

\subsection{Implications for the Nature of \agn}

\agn\ is a Seyfert galaxy that shows strong variability of optical
continuum and lines over timescales of months and longer
(see, e.g., \citealt{SM1990}).  Our work suggests that it also shows
a strong variability in its X-ray flux over timescales of months
and years.  The variability level seen in Figure\,\ref{agnlong}
is comparable to that seen in the well-studied Seyfert galaxies
such as NGC~4051 (see, e.g., \citealt{Pea2000}), IRAS~13224$-$3809
\citep{Boea1997} or Akn~564 \citep{Tea2001}.

\agn\ also shows strong, shorter-term X-ray variability,
by a factor of 3 in 8 hours (\citealt{Kea1997}; see also \citealt{T1980}).
Our new \xte\ data also show strong variability on time scales of days,
indicating such a behavior is quite common.  Given this rapid variability
in the optical and X-ray bands, and the relatively modest width of the
emission lines, we suggest that it probably belongs to the class of
Narrow Line Seyfert 1 (NLS1) galaxies (for an overview of X-ray properties
of these objects and correlations of X-ray and optical properties, see,
e.g., \citealt{Boea1996}).  Since the emission lines are strongly variable,
this source could have escaped proper identification, depending on its
state when the census was taken.  NLS1s often have complex X-ray spectra,
with a hard X-ray continuum above $\sim$1 keV and a strong soft excess
below, so it would be interesting to study its spectrum with an X-ray
spectrometer with high and low energy sensitivities, such as {\sl XMM-Newton}.

\section{Results on \cv}

\subsection{Optical Photometry}

The main features of the optical light curve of \cv\ were described by
\cite{Pea1995}, which contains details of the 1993 and 1994
photometry campaigns.  We have continued this coverage steadily
through 2002, including a 140 day campaign in 1998 to support the
\xte\ observations, using primarily the telescopes of the Center for
Backyard Astrophysics (CBA; \citealt{SP1993}).
Details of these follow-up campaigns will be reported elsewhere.
We make use of two key findings from the optical campaigns in
interpreting the X-ray data:

(1) The timings of the optical eclipses can be described by the following
linear ephemeris in Heliocentric Julian Date on the UTC system:

\begin{equation}
Mid-eclipse=HJD 2449199.693+0.14023475(5) E
\end{equation}

This ephemeris is used throughout this paper in studying the eclipse
and other orbital modulations.

(2) The spin period has been steadily decreasing through 1993--2002.
We have measured the times of primary spin minima from light curves averaged
over the 50-day beat cycle (to avoid contamination from the orbital
variability), and report the results in Table\,\ref{spinmin}.
We plot the observed minus that calculated (O$-$C) times using a test
linear ephemeris, $2449197.741+0.140613 E$ (Figure\,\ref{spinoc}).
The parabola represents our best-fit quadratic spin ephemeris:

\begin{equation}
Spin-minimum=HJD 2449197.741 + 0.140630 E - 6.5\times10^{-10}E^2
\end{equation}

This equates to a period decrease with $\dot P = -(9 \pm 1) \times 10^{-9}$,
confirming and extending the result of \cite{GS1997}.  Their recent update
of this result \citep{Sea2003} is also compatible with this level of spin-up.
Such a rapid spin up shows that \cv\ is currently well away from spin
equilibrium, thus strongly favoring the asynchronous polar model of \cv.
This implies a synchronization time of 120$\pm$15 years, similar
to that seen in other asynchronous polars.  The beat period between
the eclipse and spin periods was 52 days during our 1998 \xte\ campaign.

\subsection{Hard X-ray Orbital Modulation}

We present the average orbital modulations of \cv\ observed with
\sax\ MECS (see Table\,\ref{obslogsa}) in the left panels of
Figure\,\ref{saxav}.  Note that the 8 \sax\ observations were
spread over the beat cycle, and therefore we might expect
features that are persistent through the beat cycle to be prominent
in this representation.  The X-ray eclipse, seen at the phase of the
optical eclipse as predicted by ephemeris (1), is deep and shows sudden
transitions, which we will analyze in detail in \S 4.3 using the \xte\ data.
In addition, two energy-dependent dips can be seen.  These dips are
deeper in the 1.5--4.5 keV band than in the 4.5--8 keV band, hence
show up as periods of enhanced hardness ratio [(4.5--8 keV)/(1.5--4.5 keV)].
The hardness exceeds 1.0 between orbital phase 0.17 and 0.24, and between 0.25
and 0.27; and between 0.56 and 0.72.  Given these ranges, we will refer
to these dips as being at orbital phase 0.2 and 0.65.  Although the average
intensity is lower immediately after the eclipse, the hardness ratio
is lower at this phase, if anything, than the rest of the orbital cycle.

The average \sax\ MECS spectrum of \cv\ is typical of magnetic CVs
in showing a hard continuum and the Fe K$\alpha$ emission feature.
The hardness of the continuum is indicative of a complex partial
covering absorber \citep{DM1998}.  Given this, and the presence
of orbital, spin, and beat periods in \cv, the existing data are
not good enough to allow a quantitative description of spectral
changes beyond what can be seen from hardness ratio variations.
In particular, although we note that the Fe K$\alpha$ feature
appears to be variable and that this potentially provides a useful
tool, it is not even clear what period the lines are modulated on.

We have also folded the same \sax\ MECS data, excluding those taken during
orbital phase 0.95--1.10, on the spin ephemeris (2) above, which we present
in the right panels of Figures\,\ref{saxav}.  While these light curves show
strong variability, a systematic pattern is not obvious; moreover, the
the spin-folded hardness ratio curve is more stable than the orbit-folded
hardness ratio curve.  We interpret this as indicating that the orbital
modulation is more important in hard ($>$1.5 keV) X-rays than the spin
modulation.  While the latter is expected to be present at some level,
we cannot claim detection of one in the \sax\ data.  This is in contrast
to the previous claims \citep{GS1997,Sea2003} of X-ray spin modulation;
note, however, these are based on \rosat\ PSPC data, and may apply only
to soft X-rays.  We also note that the \sax\ observations are particularly
well suited to discriminate between the orbital and spin phenomena in \cv.
With a lesser coverage (typically lasting 0.5--3 days, much less than
the beat period), an object which is modulated purely on the orbital period
would show an apparent modulation when folded on the spin period.  It is the
combination of the 8 \sax\ observations, spread through a beat cycle, that
enables us to show the relative importance of the orbital modulation.

Are the dips at orbital phases 0.2 and 0.65 persistent?  To answer this
question, we plot in Figure\,\ref{saxeight} the folded hardness ratio curves,
which appear to be less affected by random variations than intensity curves,
for each of the 8 \sax\ observations.  Since each of
the 8 observations was taken in 5--7 hrs, much shorter than the 50-day beat
period, the orbital and spin folds are indistinguishable, apart from a
constant phase offset.  This offset changes from observation to observation;
to guide the eye, we have indicated the location of spin phase 0.0 in each
panel of Figure\,\ref{saxeight}.  Any feature fixed in the spin period
of \cv\ should move to keep its position fixed relative to these vertical
marks; we find no such feature.  The phase 0.65 feature seen in
Figure\,\ref{saxav} is clearly detected in observations 2, 3, and 4,
each within orbital phase range 0.6--0.7.  It is not strongly detected
in observations 1 or 6, while the remaining 3 observations did not cover
the relevant phase.  The coverage of orbital phase 0.2 was poorer, but
this feature clearly varied in strength between observations 1, 2, and 3.

In Figure\,\ref{asca}, we present the folded \asca\ GIS light curves in
two energy bands, and their softness ratio (we chose this representation
as the 0.7--2 keV count rates are consistent with zero at many
orbital phases).  This observation had a total duration of about 28 hrs,
and hence this orbital fold can also be considered a spin fold with an offset.
Although at lower quality, we confirm the total eclipse in the 2--10 keV
band coincident with the optical eclipse.  The behavior of \cv\ during
this observation was similar to that it displayed during the \ginga\ 
observation (see Figure 13 of \citealt{Wea1995}).  While the \asca\ data
show a minimum at the calculated phase of the optical spin minimum (shown as
vertical lines), it is also at orbital phase 0.63.  From \asca\ data alone,
we cannot tell if this is an orbital or a spin feature; however, given
the persistence of orbital phase 0.65 feature in the \sax\ data, an orbital
interpretation seems likely; the other minimum at orbital phase 0.2 is also
reminiscent of the feature in Figure\,\ref{saxav}.

\subsection{The Eclipse}

We plot in Figure\,\ref{avecl} the average eclipse profiles of \cv\ as
observed with \sax\ MECS and with \xte\ PCA, folded on ephemeris (1).
They both show that X-ray eclipse is coincident with the optical eclipse,
and has a flat-bottomed profile typical of total eclipse by the secondary.
MECS count rate drops to $(1.0\pm 1.4) \times 10^{-3}$ \cps\ (mean and
standard deviation) during the flat bottom, compared to
$(6.0\pm 2.6) \times 10^{-2}$ \cps\ out-of-eclipse: this is a total eclipse
with no detectable residual flux
in the MECS 1.5-8 keV band.  Examination of the energy-resolved eclipse
curve is consistent with this, in particular there is no detectable
residual flux in the harder (4.5--8 keV) band.  The \xte\ PCA curve
bottoms out at $0.12 \pm 0.13$ \cps.  Since any residual flux at this
level can be interpreted as due to contamination by \agn\ (see \S 3.1),
our \xte\ result is also consistent with a total eclipse in \cv.
Given this, we believe that the ``absorption dip'' interpretation
of \cite{Wea1995} is no longer tenable.  We therefore proceed
further with the interpretation that \cv\ shows a true eclipse
by the companion star.

Visual inspection of the individual eclipse light curves shows that
the ingress and the egress are usually very sharp, and can be located
accurate to $\sim$2 s by eye (cf. Figure\,\ref{xteagn}).  However,
the out-of-eclipse flux is too low in some pointings to provide reliable
estimates (that is, the limiting factor appears to be the out-of-eclipse
flux level, not the reliability of the particular method used to measure
the timings).  We have therefore attempted to measure
the eclipse timings by eye, and report the result in Table\,\ref{xteecl},
together with the inferred eclipse widths. Blanks indicate that no
measurements were possible, while colons indicate uncertain results.

We present this in a graphical form in Figure\,\ref{indecl}, together
with the average count rates during orbital phase 0.04--0.09 (``post-eclipse
flux'').  The flux shows a clear modulation; moreover, the shape of the
modulation suggest it has a $\sim$50 day period, indicative of a beat cycle
phenomenon.  As to the eclipse transition timings, there are instances of
ingress and/or egress shifting earlier in phase over several weeks, with
sudden jumps to a later phase, but it appears to have irregular variations
as well.

We have then constructed two composite eclipse profiles.  In one,
we selected light curves for which we have been able to measure
the ingress timings; we shift each curve to align the ingress
before averaging the light curves.  The other composite profile has the
egress aligned.  These are shown in Figure\,\ref{alignedecl}.
These composite profile show sharp eclipse transitions,
that ingress and egress times do not vary in concert, and that
there may be two distinct emission regions.

The first point is obvious in the expanded views on the right half of
Figure\,\ref{alignedecl}: the composite ingress is about 2.5 bins, or
5 s, while the composite egress is slightly longer (6--7 s).  These
numbers may be overestimates since our measurement errors can only blur
the composite transitions.  Although we cannot be sure that eclipse
transitions are {\sl always\/} sharp, it is clear that they often are.
The sharpness also validates our measurement
technique, and suggest that there is an intrinsic jitter in the accretion
region location to cause the noisy appearance of Figure\,\ref{indecl}.
The fact that the egress is gradual in the ingress-aligned composite
profile (and vice-versa) shows that the two transitions do not move
in step.  Unless the accretion spot moves substantially within $\sim$ 700 s,
this means that the eclipse width also varies.  This interpretation is
confirmed by Figure\,\ref{wide}, in which we compare the average eclipse
profiles of observations 1, 2, 13, and 14 (taken during days 0--5 and
50--55 of our campaign), against that of observations 3, 4, 5, 15, 16, 17
and 18 (days 8--16 and 58--66).  The latter group has weaker out-of-eclipse
emission (the part of the beat cycle during which the post-eclipse flux
is steeply declining; see Figure\,\ref{indecl}), and hence it is difficult
to measure the individual eclipses (note the blank and uncertain entries
for these eclipses in Table\,\ref{xteecl}).  However, when combined, it becomes
clear that there was an eclipse which was noticeably wider ($\sim$750 s)
than average.  This can be explained if the location of the accretion
region moves in latitude as well as in longitude (see \S 5.4).  Finally,
the egress-aligned composite (Figure\,\ref{alignedecl}) shows a slight
rise, beginning about 0.01 cycle before the sharp egress, suggesting
that a part of the X-ray emission region becomes visible about 120 s
before the main emission region.  However, this may well be due to a
subset of the eclipses.

\section{Discussion}

\subsection{Confirmation of the Eclipsing Nature of \cv}

We have found that the X-ray eclipse is total, as observed with the
imaging \sax\ MECS and \asca\ GIS detectors.  Therefore we have no
doubt that this is not a dip due to accretion stream, but represents
a true eclipse by the secondary, as \cite{Pea1995} argued (see also
\citealt{GS1999}).  This is contrary to the conclusion that \cite{Wea1995}
reached, so it is important to examine where our disagreement originates.

\citet{Wea1995} have listed two reasons for believing that this was an
accretion stream dip.  One is that the optical eclipse light curve
of \cv\ was unlike that of other eclipsing polars.  We agree with
this assessment, and interpret this simply as meaning \cv\ is unlike
other polars, perhaps related to its asynchronous nature.  The other
reason is what they believed to be a residual flux from \cv\ during
the eclipse, which was harder than the out-of-eclipse flux.  Given
that they based this on a non-imaging \ginga\ LAC observation pointed
at \agn\ (see Figure\,\ref{map}), and given that our far more extensive
X-ray data (both imaging and non-imaging) contradict this, we believe
that the residual flux was in fact from \agn\ or yet another source in
the LAC field of view.  We believe that past non-imaging X-ray observations,
particularly those which had \agn\ on-axis, are liable to be contaminated
by this highly variable Seyfert galaxy, and cannot be re-interpreted as
solely due to X-rays from \cv. At some epochs, as during the
\rosat\ All-Sky Survey (Figure\,\ref{map}), the AGN is brighter than the CV.

Our conviction is strengthened by the relative stability of the feature.
A simple folding (Figure\,\ref{avecl}) results in a well-defined and sharp
eclipse profile, with the only ill effect being the blurring of the
transitions.  This is consistent with an eclipse by
the secondary of a small accretion region (or regions) moving about on
the white dwarf surface; we do not believe that a stream dip in an
asynchronous polar can be stable to this degree.

Given that the X-ray eclipse is by the secondary, we must also accept
that the optical \citep{Wea1995} and UV \citep{SS2001} eclipse is also
by the secondary.  This implies that the geometrical pattern of
optical emission is unusual among polars, and that the UV continuum
is not entirely due to the white dwarf photosphere, albeit with
a superficial similarity.  The accretion stream is a likely candidate
for the additional source of optical and UV emission; this possibility
should be investigated further using optical and/or UV data.

\subsection{Inclination and the White Dwarf Mass}

The eclipse width, customarily defined as time between mid-ingress
and mid-egress, defines the relationship between the mass ratio $q$
and the binary inclination $i$, if the eclipsed body is centered on
the center of the white dwarf.  Although this last assumption may be
incorrect by up to the white dwarf radius for the X-ray emission region
in \cv, an eclipse width of 695 s is indicative of $i \sim 75.7^\circ$
for $q$=0.5 and $i \sim 77.7^\circ$ for $q$=0.35 (Figure\,\ref{incmass}).
In constructing this plot, we have used the Roche-lobe shape, as 
in \cite{B1979}.

Further, we take the 85 s egress duration in the average \xte\ profile
(Figure\,\ref{avecl}) to be the result of the accretion region shifting
on the white dwarf surface.  This can be translated into a minimum size
of the white dwarf, hence its maximum mass, using a white dwarf mass-radius
relationship\footnote{We have used the convenient formula in \cite{PW1975}.
Use of alternative formulae, or the effects of different composition or core
temperature, do not affect our conclusions significantly.}.
In the upper panel of Figure\,\ref{incmass}, we plot the full
duration of the egress as a function of $q$, for two assumed values of mass
of the secondary.  In this and subsequent calculations, we have replaced
the Roche lobe with a sphere that exactly reproduces the $q$--$i$
relationship, given the 695 s eclipse duration, to speed up the calculation.
The solid line is for the secondary mass M$_2$ = 0.31 M$_\odot$, corresponding
to a normal main sequence mass-radius relationship.  For the maximum
theoretical duration not to exceed the observed duration of 85 s,
$q$ must be less than 0.52, or the primary mass M$_1 <$ 0.59 M$_\odot$.
For a 0.22 M$_\odot$ secondary (its radius R$_2$ of 1.2 $\times$ a main
sequence star of the same mass), then the $q$ limit shifts considerably,
but the limit for M$_1$ remains about the same.  Thus we conclude that
the white dwarf in \cv\ is not particularly massive.

Although this conclusion is not warranted if the X-ray emission arises
high above the white dwarf surface (thus allowing the location to move
by much more than one white dwarf diameter), this is difficult to reconcile
with the compact size of the X-ray emission region implied by the rapid
eclipse transitions (Figure\,\ref{alignedecl}).  The only way around this
dilemma would be a vertically compact region high above the white dwarf
surface (i.e., with little X-ray emission lower down); however, we consider
this rather unlikely on theoretical grounds.  If we take the combined
durations of ingress and egress  in the aligned profile
(Figure\,\ref{alignedecl}) of 12 s to be the maximum contribution from
the vertical extent of the emission region,
and subtract it from 85 s, we still require a large, low-mass primary
with an estimated upper limit of 0.67 M$_\odot$.  This contradicts
the estimate of 0.98 M$_\odot$ (90\% confidence range of 0.78--1.19
M$_\odot$), based on the partial covering cold absorber fit
to the the \xte\ spectrum of \cv\ \citep{R2000}, and implies the presence
of hitherto unaccounted-for source(s) of systematic errors in either
or both methods of X-ray based mass determination.

The wider eclipse sometimes seen can be most easily explained if
the location of the observed X-ray emission shifted in latitude.
This could be caused either by accretion switching from one pole
to the other, or, if both poles are accreting all the time, by one
pole going behind the white dwarf limb and the other coming into view.
The required shift to widen the eclipse from 700 s to 750 s is of
the order of 6000 km, comparable to the radius of the white dwarf (R$_1$).
Such a large shift suggests that the dominant source of X-rays
is associated with the magnetic pole on the upper hemisphere
(when the eclipse width is $\sim$700 s) and with the lower pole
when the width is wider (see \S 5.4 for further discussion). 
If this is true, then the above binary inclinations are somewhat
(by a few 10th of a degree) underestimated.

\subsection{Size of the Accretion Spot}

How many accretion spots do we see?  Whereas multiple spots cannot
be ruled out, we do not have any clear instances of two sharp ingresses
or two sharp egresses.  That is, the X-rays observed near orbital phase 0.0
appear to be always dominated by a single spot.  However, the eclipse
experiment can only find an accretion spot on the hemisphere facing
the secondary.  We infer the presence of at least one more accretion
spot from the fact that \cv\ does not have a faint phase in which hard
X-rays are essentially undetected, as observed in several polars.

The ingress and egress durations in the aligned profiles
(Figure\,\ref{alignedecl})
constrain the size of the dominant accretion spot facing the secondary.
In the directions perpendicular to the secondary's edge at ingress and
egress, the spot size is roughly 1100 km by 1600 km for a 0.62 M$_\odot$
primary and a 0.31 M$_\odot$ secondary (or 0.13 R$_1$ by 0.19 R$_1$);
the numbers change to 1300 km by 1900 km (0.20 by 0.29) for a 0.87M$_\odot$
primary.  This is much larger than found in normal,
phase-locked polars.  For example, in HU~Aqr, \cite{Sea2001}
report an eclipse egress of only 1.3 s, corresponding to 450 km.
In the only X-ray eclipsing IP, XY~Ari, \cite{H1997} found an 
egress of 2 s.

The above values concur with the traditional picture that
normal polars should have the smallest spots, since they accrete from a
stream with a small cross-sectional area.  It has previously been 
supposed that IP accretion areas will be much bigger, since
the partial accretion disk might feed accretion over a much
larger range of azimuth.  However, the XY~Ari value is much
close to that for HU~Aqr than that for \cv.  This can be explained
if the stream in polars punches into the magnetopshere and is stripped
over a range of radii, while disk-fed accretion in IPs is
from a smaller range of radii, compensating for any larger range of
azimuth \citep{H1999}.

In \cv, both effects could be combining to produce the
much larger accretion area.  It is stream fed, leading
to a range of stripping radii, but the accretion flow also
has a large azimuthal extent (\S 5.5).  The accretion may switch
from pole to pole, and this non-stable nature could contribute
to the disorganized state of the accretion, further increasing the
accretion footprint.  However, if this is the case, we might expect
the larger spot size to lead to lower temperature of the soft component.
In contrast, \cite{Fea1996} estimates a high temperature (kT=60$\pm$35 eV
for the blackbody component in a bremsstrahlung plus blackbody fit)
from the \rosat\ PSPC spectrum of \cv, compared with a typical
PSPC-measured temperature of $\sim$30 eV
\citep{Rea1994}.  Although these are not necessarily inconsistent
(soft X-rays could originate in a small region within the larger
envelop defined by the hard X-ray eclipse light curves), further
studies are necessary to clarify the situation.

\subsection{Movement of the Accretion Spot}

The gradual shifting-forward of ingress/egress timings over $\sim$20 day
period suggests that the accretion region is moving retrogradely
in the co-rotating frame of the binary.  Moreover, the fundamental period
indeed appears to be $\sim$50 days.  These are consistent with the
asynchronous polar model of \cv, and are inconsistent with two of the three
versions of the IP model of \cite{M1998}.  Although one version of
the IP model is consistent with our data, the confirmation of the $\dot P$
of the 12,150 s period strongly favors the asynchronous polar model.
The inferred synchronization timescale of $\sim$ 120 yrs is comparable to
that for V1500~Cyg \citep{Sea1995}, while we naively expected a clear
difference for \cv\ in which accretion and synchronization torques
cooperates (unlike in other asynchronous Polars in which they compete;
\citealt{M1998}).  Clearly, we still have much to learn about the
synchronization torques in asynchronous polars.  Moreover, if all
asynchronous polars are recent novae, such a rapid synchronization
timescale would suggest a short ($\sim$ 2000 yrs) nova recurrence time
for polars, which in turn leads to some difficulties \citep{W2002}.

Given the high inclination of \cv, it is rather unlikely that a
single accretion spot can remain in view throughout a spin cycle.
For a rotational co-latitude of the spot $\beta'$, this requires
$i + \beta' < 90^\circ$; since $i \ga 75^\circ$, it requires
$\beta' \la 15^\circ$ (cf. the study of optical spin minima
by \citealt{Sea2003}).  Any accretion spots having a $\beta'$
greater than $\sim 15^\circ$ must disappear behind the white dwarf
limb; the relative lack of spin modulation in the hard X-rays
(Figure\,\ref{saxav}) suggests that this is compensated for by
the appearance of the other pole.  A lack of hard X-ray spin
modulation in \cv\ is not a total surprise, since there is a precedent
in the flaring state of the well-known asynchronous polar, BY~Cam
\citep{Iea1991}.

In Figure\,\ref{schema}, we schematically plot the movement of these
accretion spots as seen from Earth at mid-eclipse.  In this plot, we
assume two diametrically opposed spots, moving in longitude but
staying constant in latitude.  This is a simplification: while the
magnetic poles are expected to move this way, the offset between the
accretion spot and the magnetic pole changes as a function of the magnetic
pole orientation \citep{M1988,GS1997}.  Because we cannot fully model this
without more observational constraints, we will use this as a crude model.
Furthermore, we assume that both spots accrete at all times, and we observe
the region that is not hidden behind the white dwarf limb.  In this schematic,
the only asymmetry between the upper and lower poles is introduced by the
viewing geometry (that is, $i \neq$ 90$^\circ$).  For this plot, we use
$q$=0.55, $i$=74.88$^\circ$, and $\beta'$=45$^\circ$.

Such a model makes a specific prediction for the X-ray eclipse timings.
To facilitate a better comparison with this model, we have re-analyzed
the \xte\ eclipse timing by first combining light curves taken at
similar beat phase.  For this purpose, we use the beat period of 52 days
appropriate for 1998, based on our ephemerides (1) and (2), and use
HJD 2450940.99 as relative beat phase 0.0\footnote{This is an arbitrary
epoch for beat phase 0.0, chosen simply because it is the time of the
first well-measured eclipse in our 1998 optical campaign, to be reported in
a future paper.}.  We then measured the ingress
and egress phases in these 8 light curves, with higher signal-to-noise
than in individual eclipses; we were able to measure all 8 X-ray ingresses
and 7 of 8 egresses.  These, and the resultant eclipse duration, are plotted
as a function of this relative beat phase in Figure\,\ref{beatfold}.
Overplotted are the prediction of the model shown in Figure\,\ref{schema},
for which beat phase 0.0 is defined as when the upper pole is most directly
facing the secondary.  We emphasize that this is not a fit; although the
phasing works approximately correctly, this fortuitous agreement nevertheless
appears to be imperfect.  The choice of $\beta'$ was adjusted by hand
to reproduce the gross features of the data, but no formal fitting was
performed.  In particular, we note that this reproduces the
fact that there is one jump to later phase in each timing, with a gradual
shift forward for the rest of the cycle.  For the ingress, this corresponds
to the jump from the disappearance of the lower pole to the left and the
appearance of the upper pole to the right.  The change from upper pole to
lower pole produces little shift, however, given this geometry, and vice
versa for the egress timing.  We believe this geometry provides a reasonable
framework for more quantitative modeling of the X-ray eclipse timings in \cv.

When two spots are observed in turn during a beat cycle, a sudden changes
in eclipse duration is predicted (Figure\,\ref{beatfold}), although our
particular model ovepredicts the amount.  In comparison, the single
spot model of \cite{GS1997} predicts a gradual shift with a total amplitude
of $< 20^\circ$.  The shift required by the data is $\sim$6000 km (\S 5.2),
or at least $\ga 40^\circ$ in latitude, depending on the primary mass.
We therefore consider it rather secure that the eclipse timing
data, which is sensitive to the projected displacement of the emitting
region, requires the presence of two hard X-ray emitting spots.
The spin modulation, on the other hand, is caused by the angular dependence
of emission.  Hard X-rays are probably emitted uniformly, whereas the
optical emission is likely beamed.  Our result, therefore, does not
necessarily contradict that of \cite{GS1997}.

We have used an arbitrary epoch to define a relative beat phase for our
data, which appears to be close (Figure\,\ref{beatfold}) to the one shown
in Figure\,\ref{schema}.  That is, at our relative beat phase 0.0, the upper
pole points roughly towards the secondary.  In contrast, \cite{GS1997}
have defined beat phase 0.0 as when the magnetic pole points most directly
at the threading region.  Using the epoch derived by \cite{Sea2003} and
our spin ephemeris (Eq. 2), our beat phase 0.0 appears to correspond to beat
phase $\sim$0.2 in the system of \cite{GS1997}.  This can be interpreted
as reflecting the average position of threading region in \cv, which in
many polars leads the secondary by a similar amount.

\subsection{The Nature of the Orbital Modulation}

There are two orbital dips in the \sax\ light curves of \cv, near
orbital phase 0.2 and 0.65 (Figure\,\ref{saxav}).  Although neither
persists throughout the beat cycle, when they are present, their
location appears roughly fixed in the orbital frame rather than in
the spin frame.  What are they, and what can they tell us about the
accretion geometry of \cv?

First, we suspect these are the causes of the complex appearances
of the power spectra, including the strong peak at about 4000 s,
of the \rosat\ data of \cv\ \citep{M1998}.  The fact that these appear
to be an orbital, rather than a spin, phenomenon is an additional evidence
against the IP interpretation.

However, the phasing of these dips are reminiscent of that of
dipping low-mass X-ray binaries (see, e.g., \citealt{Sea1992}).
If this analogy is real, then \cv\ may have a ring of material
circulating around the white dwarf, something akin to a partial
accretion disk.  In this picture, the accretion stream from the
secondary hits the ring, creating the dip-causing structure at
phase 0.65.  The splashed material hits the ring again at phase
0.2, causing a secondary dip.  Regardless of the validity of such
a picture, it does appear necessary for an accretion stream to go
around the white dwarf 80\% of the way before being captured by the
magnetic field.  Similar inference has also been made from optical
emission line observations \citep{Fea2000}.  We further speculate
that such a ring might be responsible in part for the unusual optical
eclipse light curves and of the optical eclipse timing variations.
It appears that we have plenty to learn about the accretion processes
in \cv\ and other asynchronous polars.

\section{Conclusions}

We draw the following conclusions from our study of \cv\ and \agn\ using
our \xte\ campaign as well as archival \asca\ and \sax\ data:

(1) \agn\ is a variable source of 2--10 keV X-rays, occasionally reaching
a few times 10$^{-11}$ ergs\,s$^{-2}$cm$^{-2}$ level.  The levels seen
during 1993--1994 with \rosat\ and \asca\ appear to be the faintest
it was ever seen.  Variability on the timescale of days and longer,
and with amplitude of a factor of a few, is well-established.

(2) Non-imaging observations of \agn\ cannot therefore be reinterpreted
as that of \cv.  These data contain unknown mixtures of X-rays from both
\cv\ and \agn, and the latter can be brighter than the former.

(3) \cv\ is an eclipsing system beyond doubt, with an eclipse
width of 695 s.  Moreover, the X-ray source is small and moves
around on the face of the white dwarf.

(4) The total range of movement suggests a low-mass white dwarf,
less than 0.59 M$_\odot$, if the X-ray emission is on the surface,
or 0.67 M$_\odot$ allowing for as much vertical extent as we think
reasonable.  These are lower than inferred from the modeling of
its X-ray spectrum.

(5) When the individual ingress is aligned before stacking the
light curves, it reveals a sharp ($\leq$ 5 s) ingress.  Ditto for egress
(6--7 s).  This suggests an accretion region, with an area of less
than 1\% of the total surface area of the white dwarf, but larger
than the spots in the IP, XY~Ari, or in the polar, HU~Aqr.

(6) The movement of the accretion spot is consistent with that expected
for an asynchronous polar in which the spin period is slightly longer
than the orbital period.  We do not have a quantitative understanding,
however, since there are currently too many free parameters.

(7) There are two orbital dips in addition to the eclipse in the
\sax\ light curves of \cv, when folded on the orbital period.
This suggests that some of the accreting matter travels most of
the way around the white dwarf before being threaded by the primary's
magnetic field.

\acknowledgments

\begin{figure}
\plotone{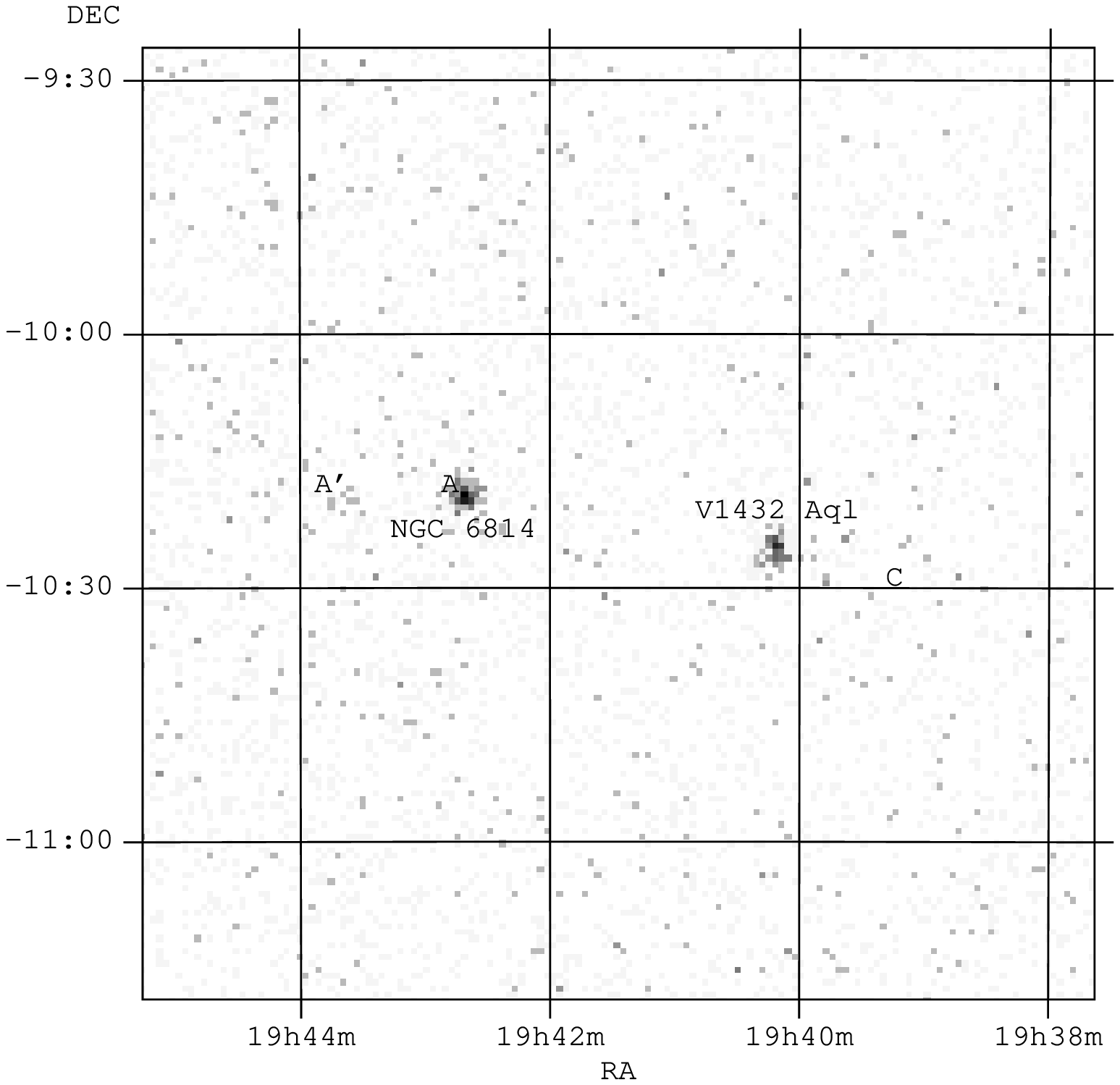}
\caption{The \rosat\ All-Sky Survey image of the region containing
\agn\ (detected at 0.36 \cps) and \cv\ (0.32 \cps), marked with
the pointing positions (A and C) used in our \xte\ observing campaign.
Also marked (A') is the intended pointing position (see text).}
\label{map}
\end{figure}

\begin{figure}
\plotone{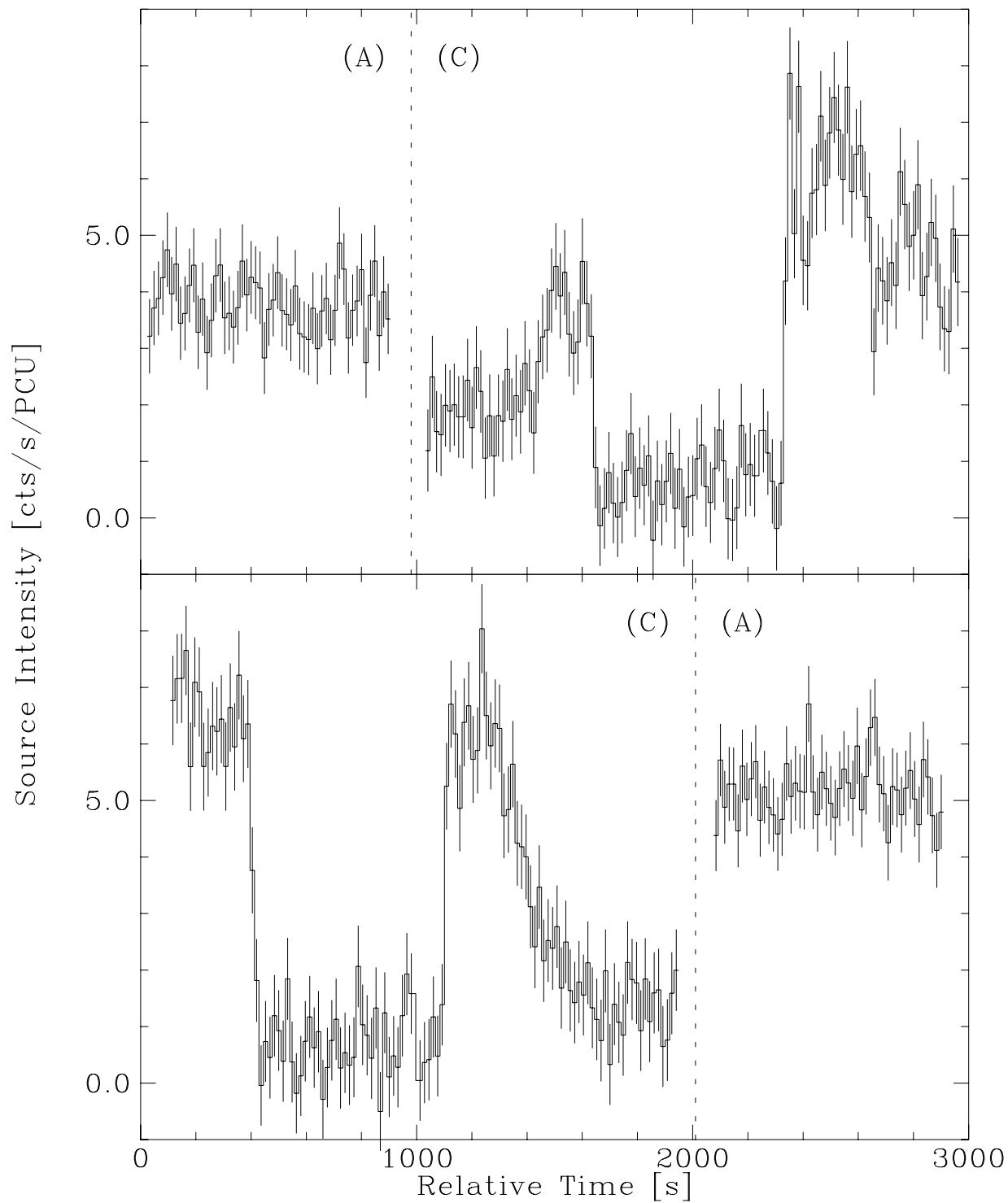}
\caption{Light curves obtained at pointing positions A and C
during the 1st and the 9th \xte\ PCA observations, likely dominated
by the \agn\ and \cv, respectively.  The bin size is 16 s,
and time 0 corresponds to 1998 May 18 at 12:58:00 UT (top)
and 1998 June 19 at 19:21:40 UT (bottom), respectively.}
\label{xteagn}
\end{figure}

\begin{figure}
\plotone{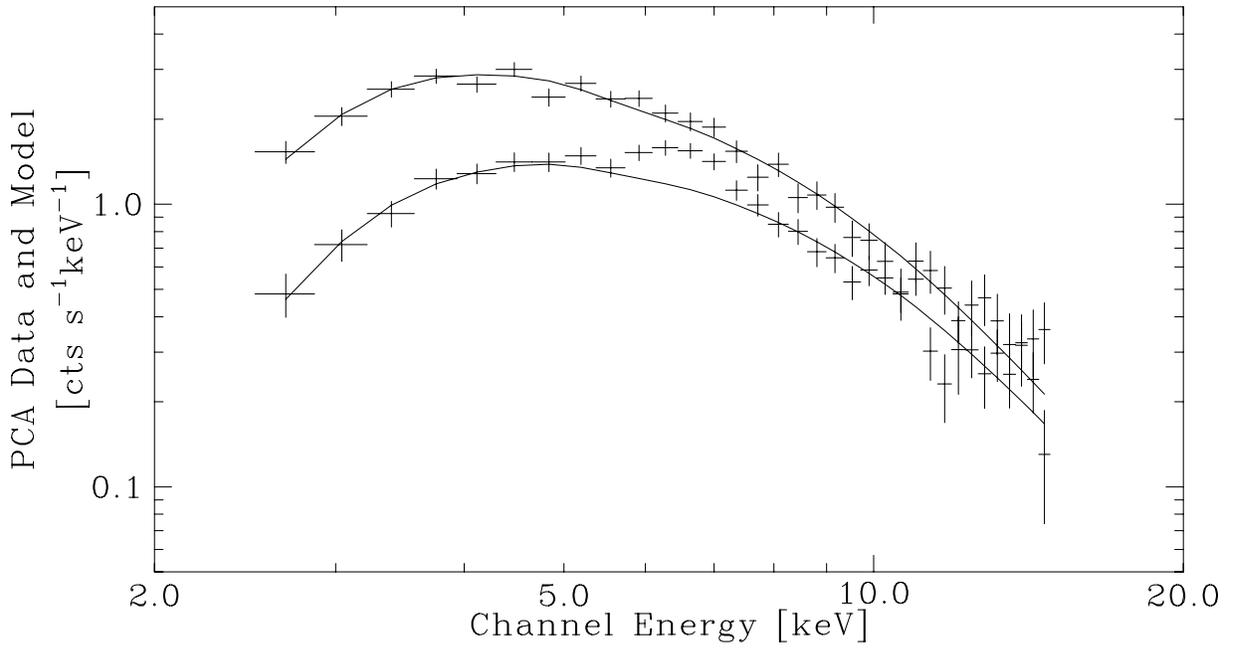}
\caption{The spectra of \agn\ (brighter) and \cv\ (fainter) from
the 1st \xte\ PCA observation.  Best-fit continuum models are
also plotted, even though an Fe K$\alpha$ is clearly required for the latter.}
\label{xteagnspec}
\end{figure}

\begin{figure}
\plotone{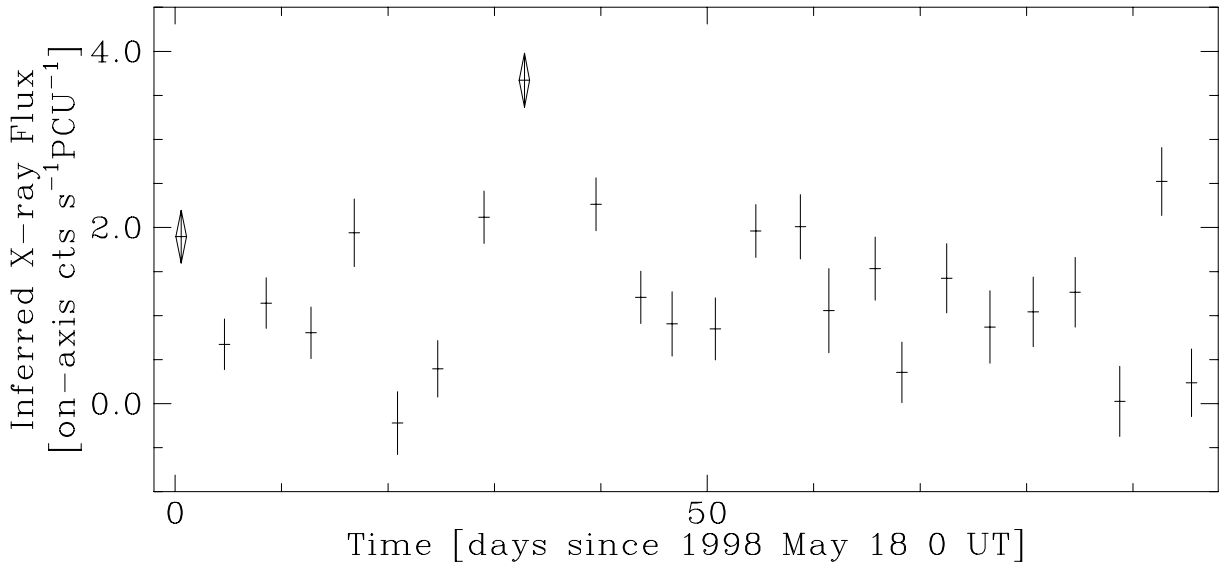}
\caption{Inferred X-ray brightness of \agn\ (appropriate for on-axis
\xte\ observations), inferred from the average PCA count rate at
mid-eclipse of \cv.  Except for the two secure
detections from ``A'' pointings (shown using combination diamond/plus
symbols),  these should be considered upper limits.}
\label{agnmon}
\end{figure}

\begin{figure}
\plotone{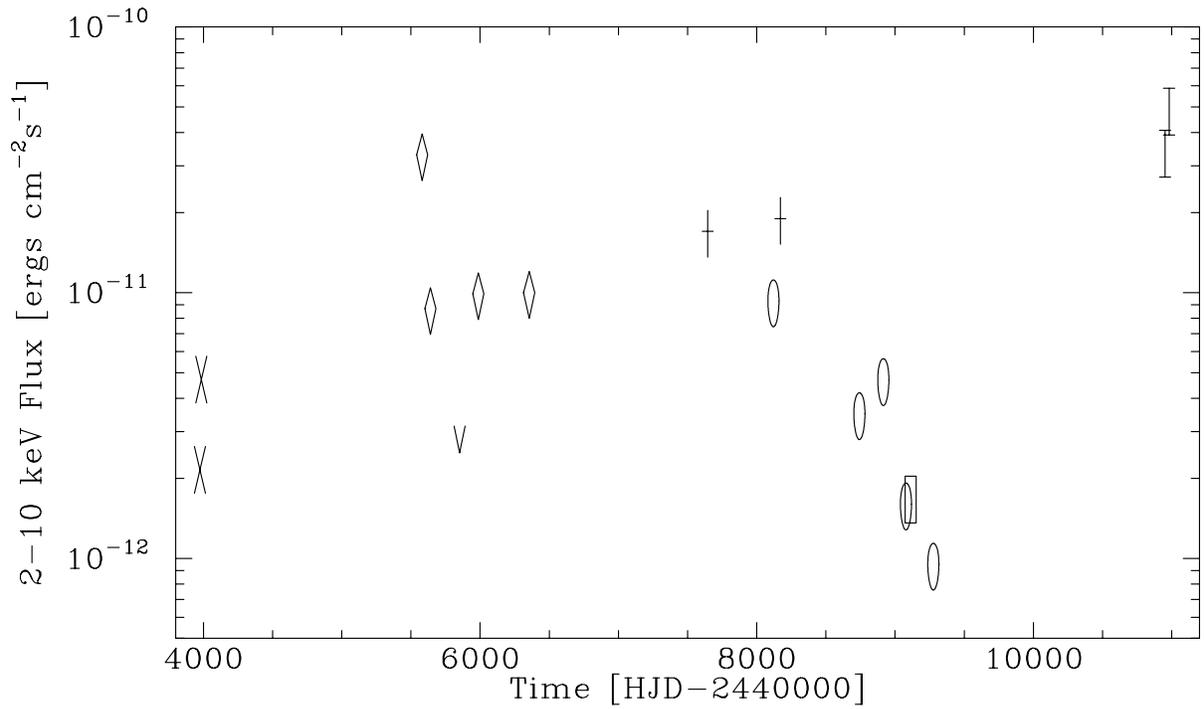}
\caption{The long-term 2--10 keV flux history of \agn\ reconstructed
from archival X-ray observations.  Symbols are: Xes for \einstein,
diamonds for \exosat\ detections and v for an \exosat\ upper limit,
crosses for \ginga, ellipses for \rosat, square
for \asca, and error bars for \xte\ observations.  See text for details
including caveats on the reliability of various points.}
\label{agnlong}
\end{figure}

\begin{figure}
\plotone{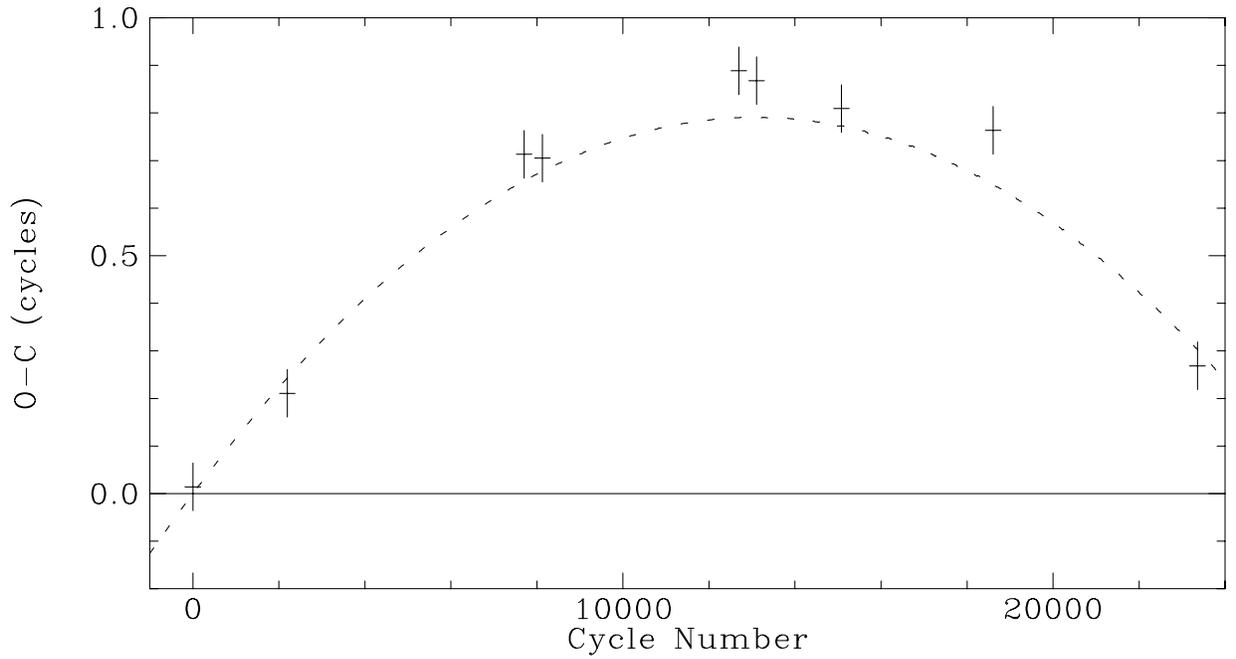}
\caption{The O$-$C diagram of the times of spin minima, measured
from CBA photometry (data given in Table\,\ref{spinmin}),
plotted as a function of cycle number relative to a test linear
ephemeris, $HJD 2449197.741 + 0.140613 E$.  Each point is the measured
times averaged over a beat cycle.  The dashed line is our best-fit quadratic
ephemeris, which corresponds to a synchronization timescale
of $\sim$120 years.}
\label{spinoc}
\end{figure}

\begin{figure}
\plotone{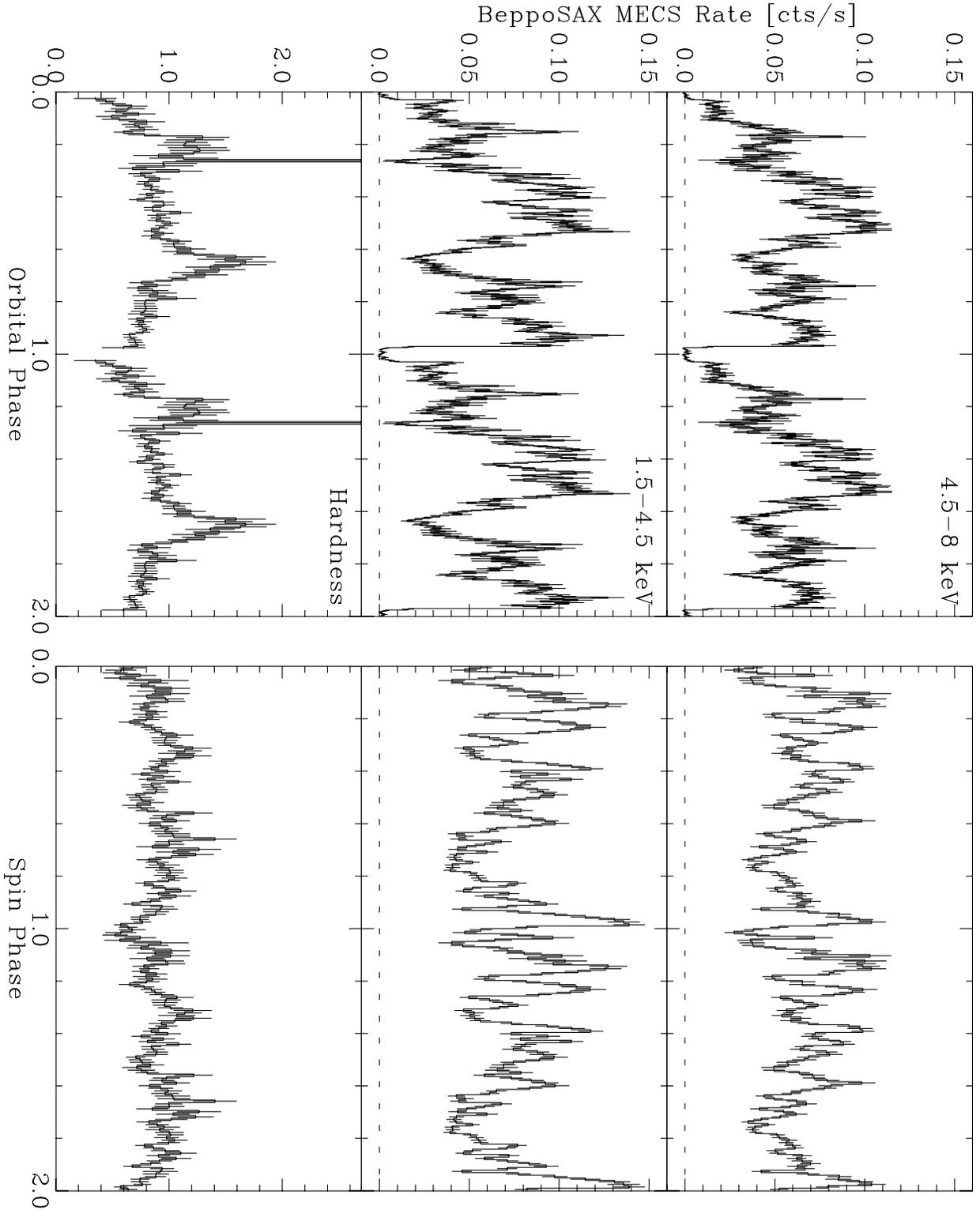}
\caption{(Left) \sax\ light curves of \cv\ in two energy bands
folded on the orbital period using ephemeris (1)
(202 bins per cycle, or $\sim$60 s per bin), and the hardness ratio
([4.5--8 keV]/[1.5--4.5 keV], 101 bins per cycle, but undefined during
eclipse), plotted twice for clarity.  (Right) The same light curves,
excluding orbital phase 0.95--1.10, folded on the spin period using
ephemeris (2) (101 bins per cycle).}
\label{saxav}
\end{figure}

\begin{figure}
\plotone{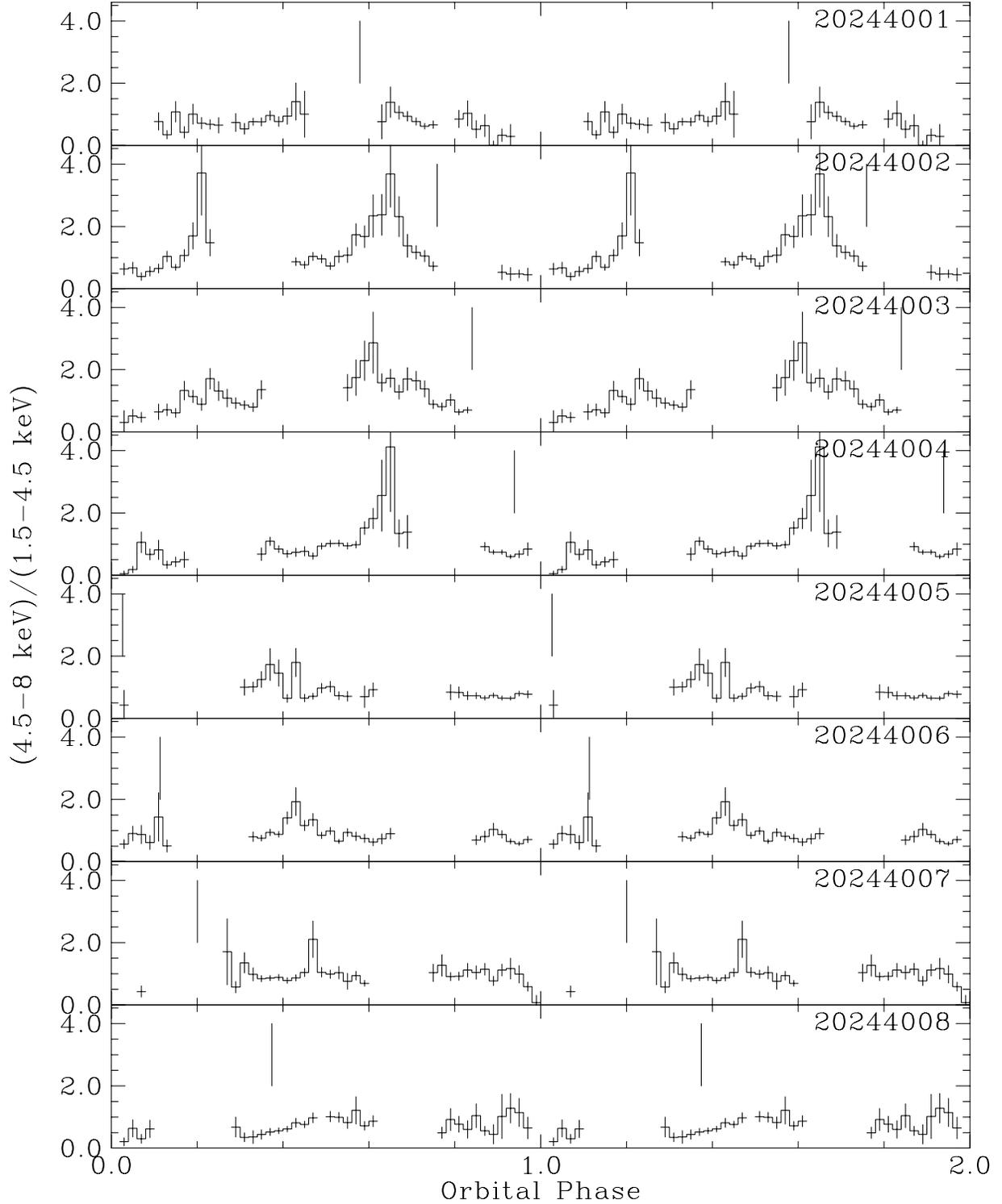}
\caption{Hardness ratios for each of the 8 \sax\ observations
folded on the orbital period (ephemeris (1); 50 bins per cycles, two cycles).
Vertical bars indicate spin phase 0.0 using ephemeris (2).
Although the data gaps complicate the comparisons, it can be seen
that spectral hardening at phases 0.2 and 0.65 is variable (see text).}
\label{saxeight}
\end{figure}

\begin{figure}
\plotone{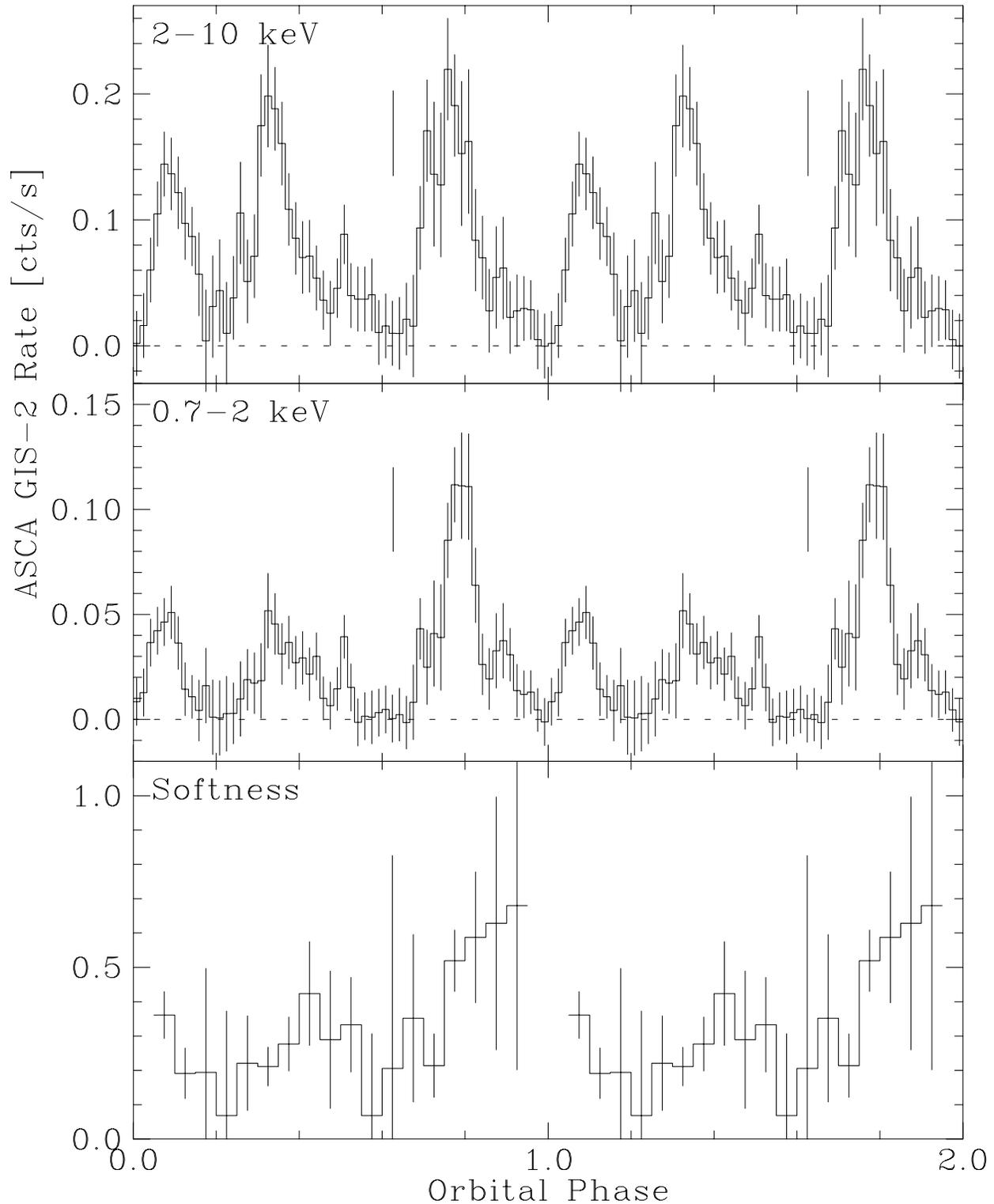}
\caption{\asca\ GIS-2 light curves in two energy bands folded
on the orbital period using ephemeris (1) (120 bins
per cycle, or $\sim$100 s per bin) and the softness ratio
([0.7--2 keV count rate]/[2-10 keV count rate], 20 bins
per cycle), plotted twice.  The vertical lines in the upper two panels
indicate the expected orbital phase of the optical spin minimum during this
observation, according to ephemeris (2).}
\label{asca}
\end{figure}

\begin{figure}
\plotone{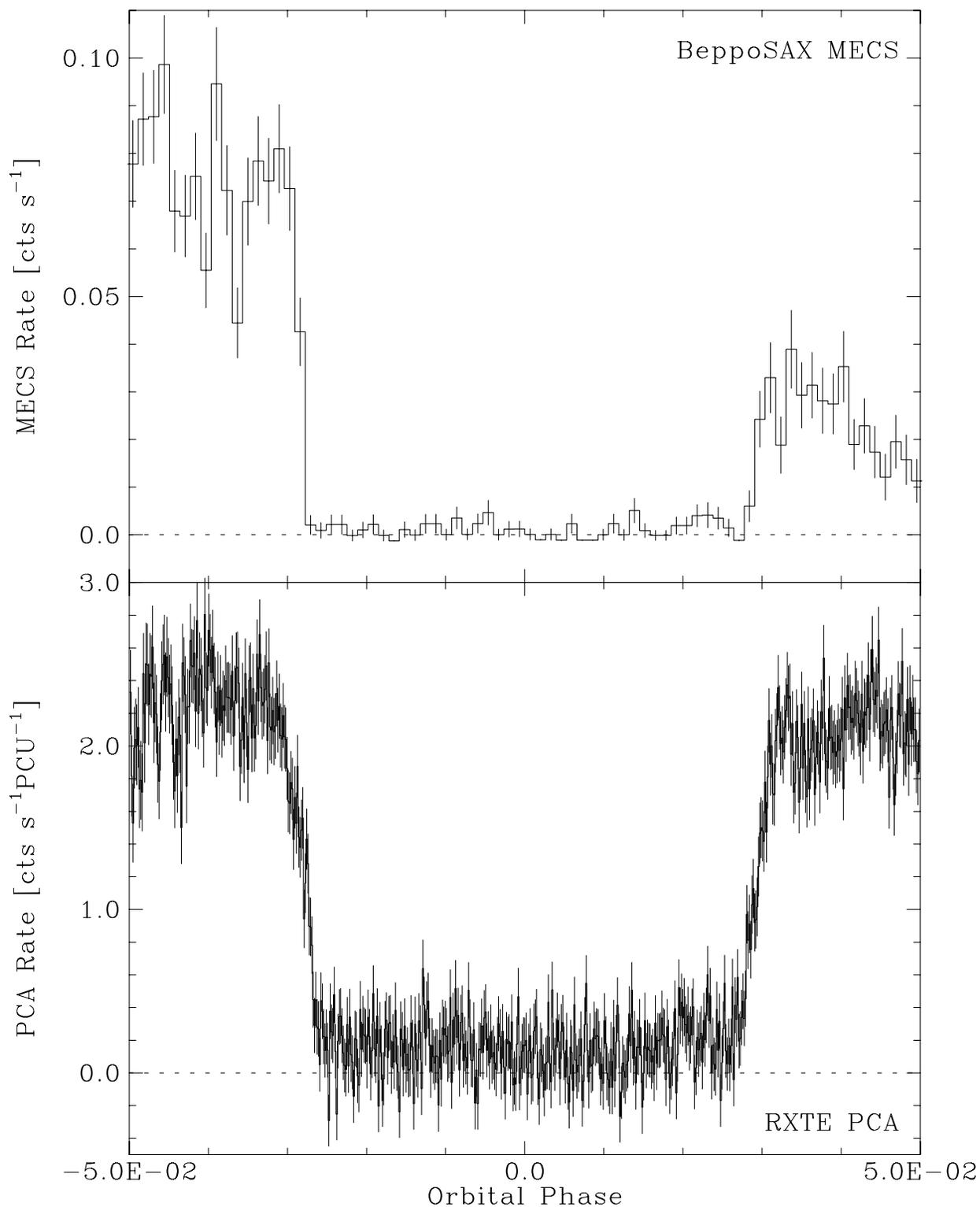}
\caption{Folded \sax\ and \xte\ light curves around phase 0.0 according
to ephemeris (1), in 757 bins per orbital cycle ($\sim$16 s per bin) for
the \sax\ data, and 6058 bins per orbital cycle ($\sim$2 s per bin) for
the \xte\ data.  Dashed lines are drawn for a net count rate of 0.0.}
\label{avecl}
\end{figure}

\begin{figure}
\plotone{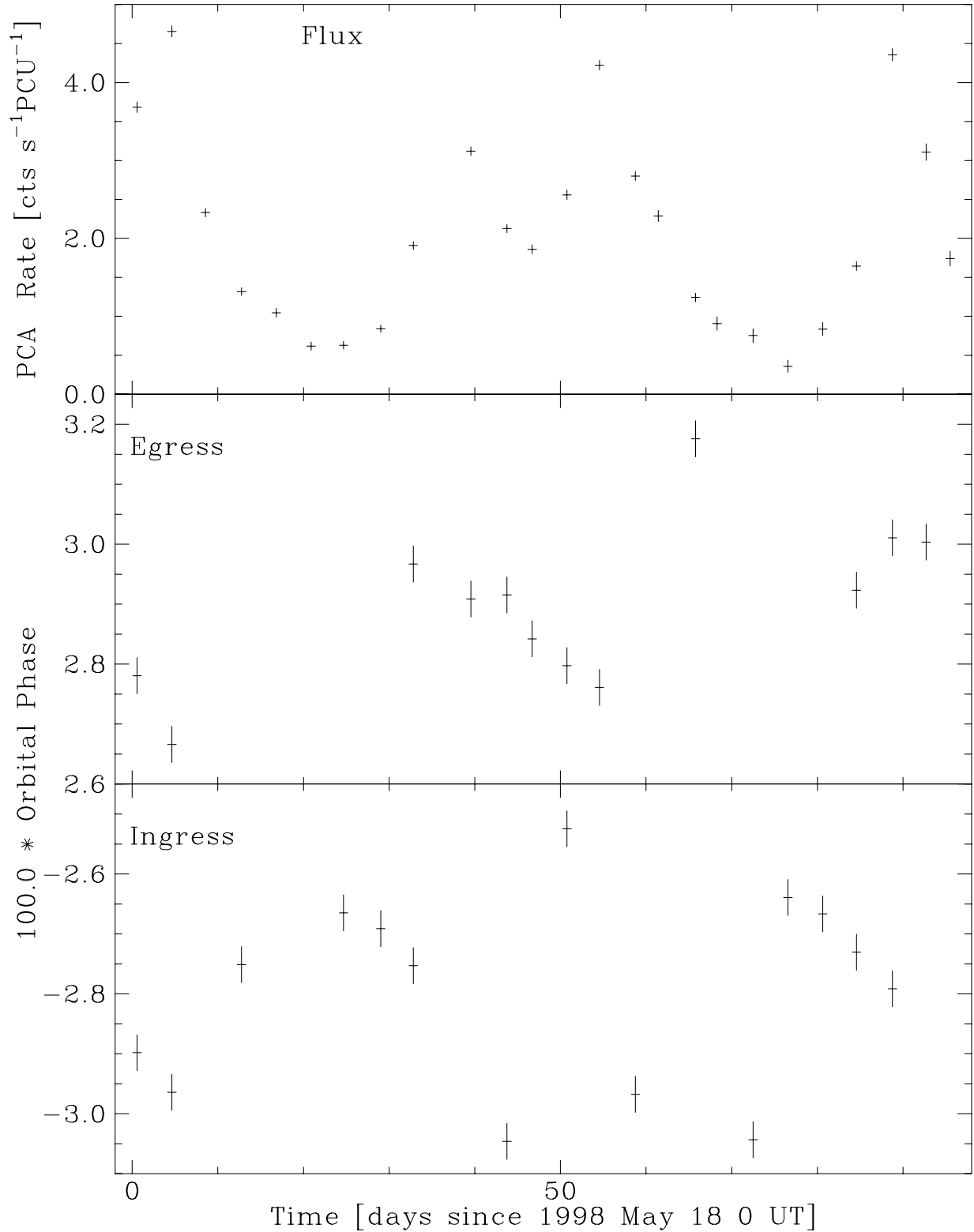}
\caption{The out-of-eclipse flux (average count rate during
orbital phase 0.04--0.09), and the ingress and egress timings
(in units of 0.01 of an orbital cycle) are plotted against time through
our \xte\ observging campaign.}
\label{indecl}
\end{figure}

\begin{figure}
\plotone{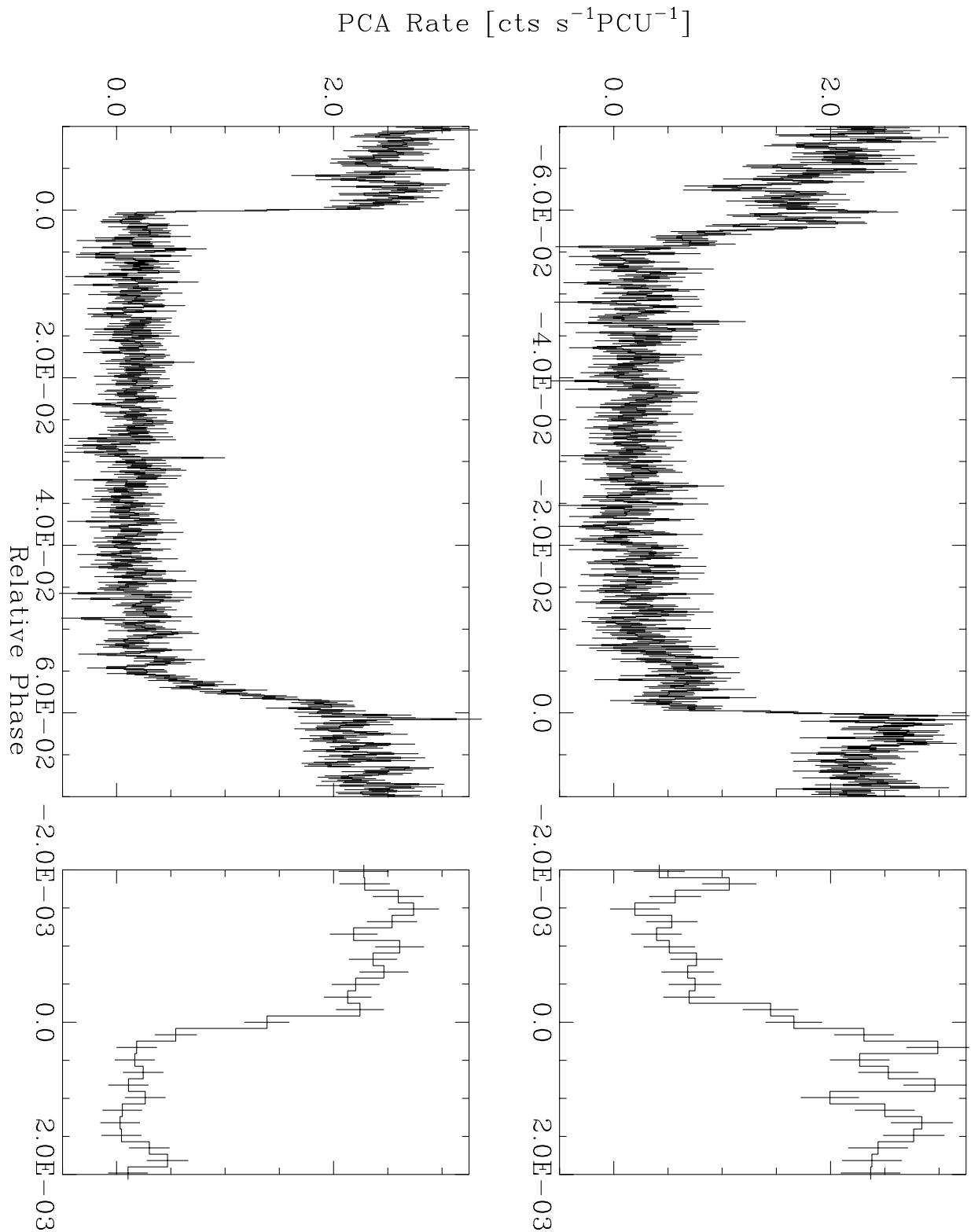}
\caption{Folded light curves in $\sim$2 s bins, after manually
aligning the ingress (lower panels) and egress (upper panels).
For this figure only, we plot phase relative to the ingress (lower panels)
or the egress (upper panels).  The left side of the figure shows the overall
eclipse shape, while detailed of the aligned feature is shown on the right.}
\label{alignedecl}
\end{figure}

\begin{figure}
\plotone{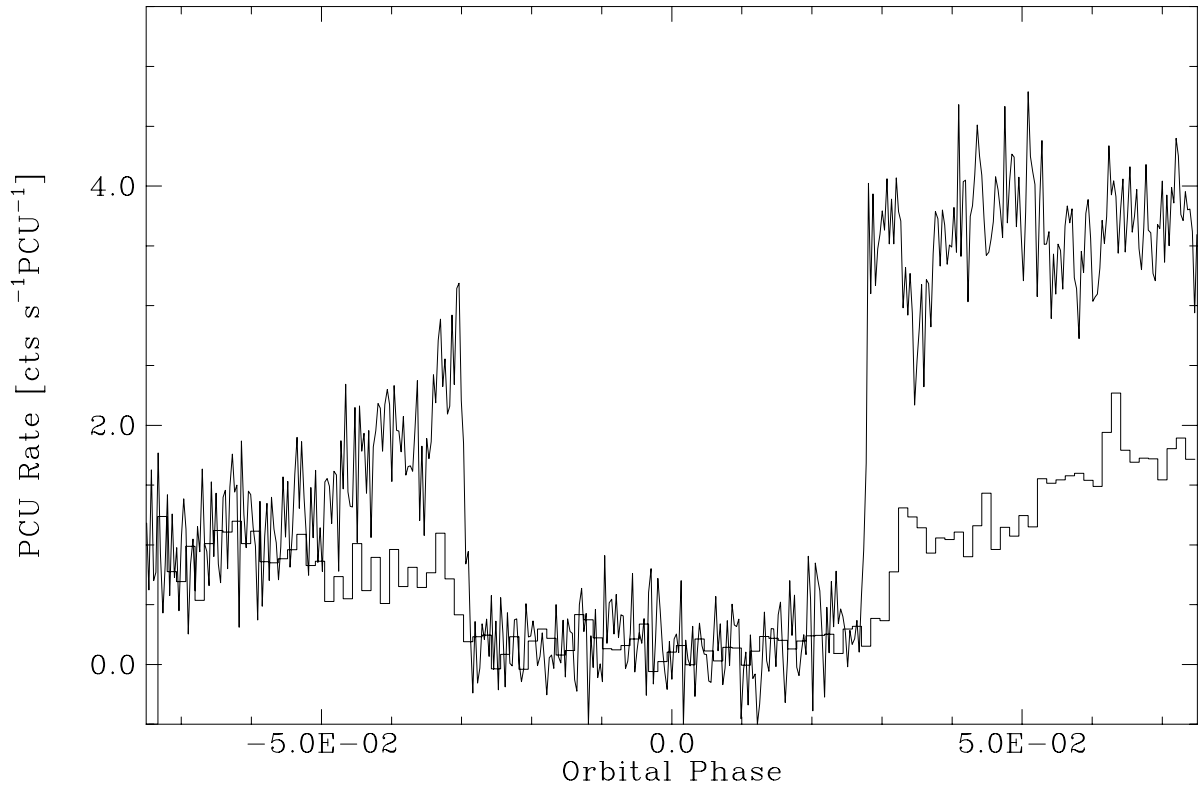}
\caption{The average eclipse profile from observations
1, 2, 13, and 14 (line plot, in 4 s per bin), overplotted
with the average eclipse profile from observations 3, 4, 5, 15, 16, 17,
and 18 (histogram, in 16 s per bin).  The latter are clearly wider
than the former.}
\label{wide}
\end{figure}

\begin{figure}
\plotone{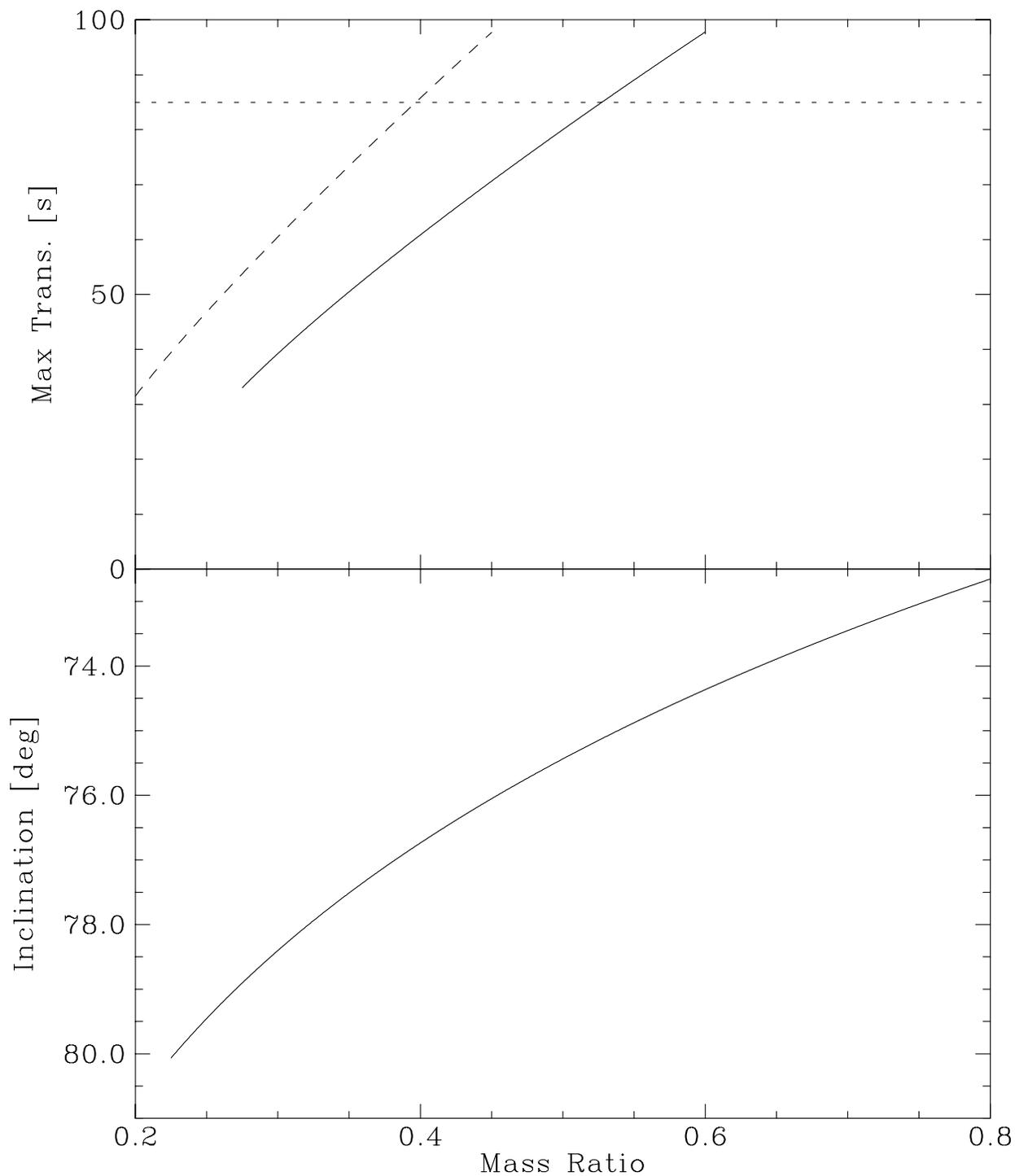}
\caption{The inclination ($i$) mass-ratio ($q$) relationship for an eclipse
width of 695 s is plotted in the lower panel.  In the upper panel, 
maximum duration of eclipse transition is plotted as a function of $q$
for two assumed secondary masses (solid line: 0.31 M$_\odot$; dashed line:
0.22 M$_\odot$).  The dotted line is the observed egress duration in the
average \xte\ data, in which the transitions are smeared due to spot
movement.}
\label{incmass}
\end{figure}

\begin{figure}
\plotone{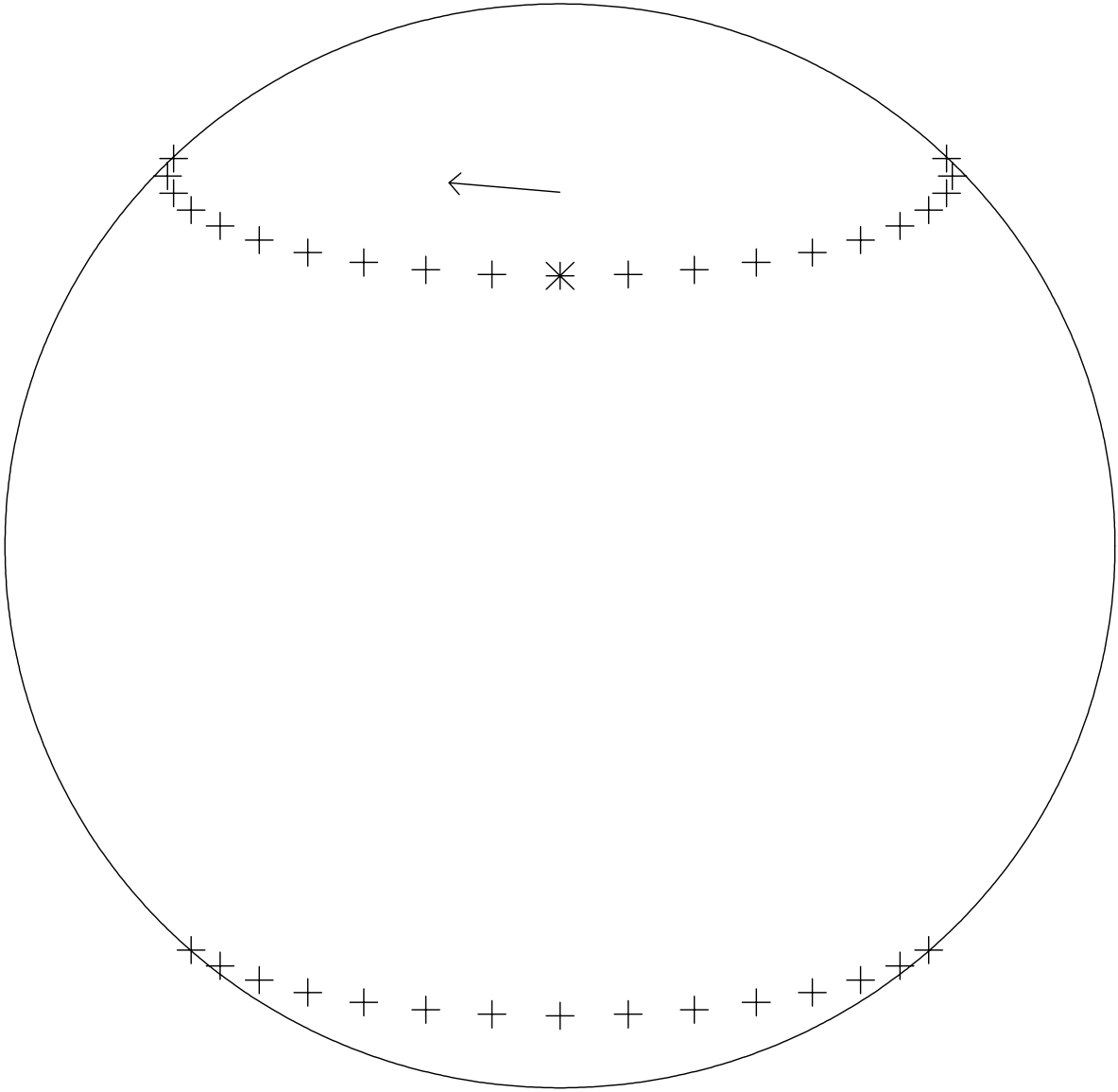}
\caption{A schematic diagram of the accretion spot movement over a
beat phase, as seen from Earth at orbital phase 0.0.  The star is the
position at an arbitrary beat phase 0.0.  The spot moves left until it
disappears and the lower pole appears to the bottom right.  It moves left
until it disappears and the upper pole reappears at top right.  The limb
of the secondary would be a line from upper left to lower right at ingress,
moving right; and a right-moving line from upper right to lower left at
egress.}
\label{schema}
\end{figure}

\begin{figure}
\plotone{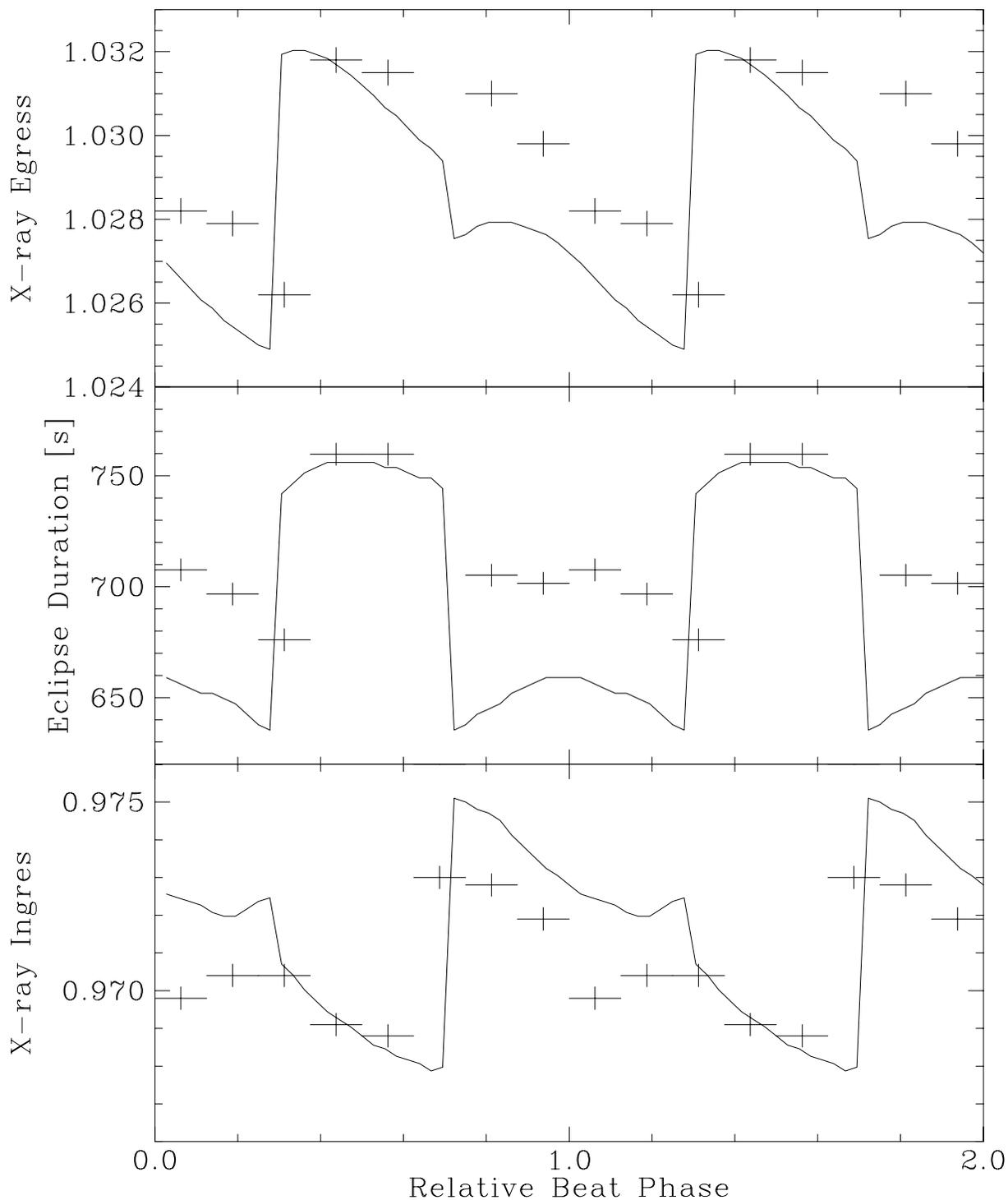}
\caption{Ingress (bottom) and egress (top) timings (expressed in orbital
phase) measured from average X-ray eclipse profiles in each of 8 beat-phase
bins, and plotted twice for clarity.  The middle panel shows
the X-ray eclipse duration in seconds.  For these points, we use a
relative beat phase defined by a period of 52 days and a phase 0.0
at HJD 2450940.99.
The lines in all three panels show the predictions of the schematic
model (see text for details).}
\label{beatfold}
\end{figure}

This research has made use of \sax\ and \asca\ data obtained from
the High Energy Astrophysics Science Archive Research Center (HEASARC),
provided by NASA's Goddard Space Flight Center.  We acknowledge the
use of NASA's {\sl SkyView\/} facility \\ (http://skyview.gsfc.nasa.gov)
located at NASA Goddard Space Flight Center in the production of
Figure\,\ref{map}.  We thank the \xte\ operations team for their
effort in scheduling these highly constrained observations, and the
PCA team for their help in deciphering the collimator response.
We also acknowledge the many observers of the Center for Backyard
Astronomy network whose observations were essential in tracking
the periods of \cv.

\clearpage

% Figure captions
%---------------------------------------------------------

\clearpage

\begin{deluxetable}{lrllr}
%\footnotesize
\tablecaption{Log of Imaging Observations}
%\tablewidth{0pt}
\tablehead{ \colhead{Satellite} & \colhead{Sequence no}
	& \colhead{Start Date \& Time} & \colhead{End Date \& Time}
	& \colhead{Exposure} \\
	 & & \multicolumn{2}{c}{(UT)} &  \colhead{(ksec)} }
\startdata
\sax  & 20244001 & 1997 Apr 02 06:50 & Apr 02 11:50 & 9.1 \\
      & 20244002 & 1997 Apr 11 16:36 & Apr 11 22:35 & 12.2 \\
      & 20244003 & 1997 Apr 15 21:42 & Apr 16 02:37 & 10.0 \\
      & 20244004 & 1997 Apr 20 20:02 & Apr 21 01:40 & 12.3 \\
      & 20244005 & 1997 Apr 25 11:10 & Apr 25 16:19 & 8.4 \\
      & 20244006 & 1997 Apr 29 19:51 & Apr 30 03:00 & 12.9 \\
      & 20244007 & 1997 May 04 10:02 & May 04 15:21 & 10.8 \\
      & 20244008 & 1997 May 13 05:44 & May 13 11:05 & 9.5 \\
      &          &                   &              &        \\
\asca & 75009000 & 1997 Oct 27 20:25 & Oct 29 00:11 & 39.6 \\
\enddata
\label{obslogsa}
\end{deluxetable}

\begin{deluxetable}{lccrr}
\tablecaption{Target and Pointing Directions of the \xte\ Campaign}
\tablehead{ & \colhead{R.A.} & \colhead{Dec} &
		\colhead{Distance to} & \colhead{Distance to} \\
	 &  \multicolumn{2}{c}{(2000)}  & \colhead{\cv} & \colhead{\agn} } 
\startdata
C    & 19 39 11.9  & $-$10 27 49.0 & 14.84$'$   & 52.00$'$   \\
\cv  & 19 40 11.47 & $-$10 25 25.1 &            & 37.16$'$   \\
\agn & 19 42 40.6  & $-$10 19 24.6 & 37.16$'$   &            \\
A    & 19 42 46.2  & $-$10 17 00.2 & 38.97$'$   &  2.77$'$   \\
\enddata
\label{obsposxte}
\end{deluxetable}

\begin{deluxetable}{rrllllll}
\tablecaption{The \xte\ Campaign}
\tablehead{ \colhead{ObsID\tablenotemark{a}} &
            \colhead{Time\tablenotemark{b}} &
            \colhead{Obs. Start\tablenotemark{c}} &
	    \colhead{End} & \colhead{Active} & \colhead{Pointings} &
            \multicolumn{2}{c}{Collimator Eff.} \\ & \colhead{(Days)} &
            \multicolumn{2}{c}{(UT)} & \colhead{PCUs} & &
	    \colhead{for CV} & \colhead{for AGN} }
\startdata
30015-01-01-00 &  0.6 & May 18 13:13 & 14:00 & 01234 & A--C & 0.8036 & 0.1133 \\
30015-01-02-00 &  4.6 & May 22 14:50 & 15:27 & 01234 & A--C & 0.8039 & 0.1133 \\
30015-01-03-00 &  8.6 & May 26 13:08 & 14:08 & 01234 & A--C & 0.8036 & 0.1132 \\
30015-01-04-00 & 12.8 & May 30 18:04 & 19:14 & 01234 & A--C & 0.8036 & 0.1131 \\
30015-01-05-00 & 16.8 & Jun 03 19:43 & 20:41 & 0123  & A--C & 0.7811 & 0.1024 \\
30015-01-06-00 & 20.9 & Jun 07 21:18 & 22:10 & 0123  & A--C & 0.7806 & 0.1022 \\
30015-01-07-00 & 24.7 & Jun 11 16:13 & 16:43 & 01234 & C--A & 0.8018 & 0.1125 \\
30015-01-08-00 & 29.0 & Jun 16 00:31 & 01:37 & 01234 & A--C & 0.8013 & 0.1126 \\
30015-01-09-00 & 32.8 & Jun 19 19:23 & 19:54 & 01234 & C--A & 0.8008 & 0.1125 \\
30015-01-10-00 & 39.5 & Jun 26 12:59 & 13:33 & 01234 & C--A & 0.7982 & 0.1123 \\
30015-01-11-00 & 43.8 & Jun 30 17:48 & 18:26 & 01234 & C--A & 0.7966 & 0.1123 \\
30015-01-12-00 & 46.7 & Jul 03 16:31 & 17:06 & 0123  & A--C & 0.7718 & 0.1010 \\
30015-01-13-00 & 50.8 & Jul 07 18:12 & 18:42 & 0123  & A--C--A & 0.7675 & 0.1027 \\
30015-01-14-00 & 54.6 & Jul 11 13:05 & 13:39 & 01234 & C--A & 0.7837 & 0.1155 \\
30015-01-15-00 & 58.8 & Jul 15 18:05 & 18:32 & 01234 & C--A & 0.7786 & 0.1011 \\
30015-01-16-00 & 61.4 & Jul 18 09:55 & 10:29 & 012   & C--A & 0.7485 & 0.0865 \\
30015-01-17-00 & 65.8 & Jul 22 18:19 & 19:24 & 01234 & A--C & 0.7560 & 0.0981 \\
30015-01-18-00 & 68.3 & Jul 25 06:49 & 07:16 & 01234 & C--A & 0.7520 & 0.0939 \\
30015-01-19-00 & 72.5 & Jul 29 11:43 & 12:14 & 01234 & A--C--A & 0.7509 & 0.0870 \\
30015-01-20-00 & 76.6 & Aug 02 13:24 & 13:51 & 01234 & C--A & 0.7514 & 0.0853 \\
30015-01-21-00 & 80.6 & Aug 06 15:01 & 15:27 & 01234 & C--A & 0.7454 & 0.0839 \\
30015-01-22-00 & 84.6 & Aug 10 13:11 & 13:47 & 01234 & C--A & 0.7429 & 0.0844 \\
30015-01-23-00 & 88.8 & Aug 14 18:10 & 18:45 & 0123\tablenotemark{d} & A--C & 0.7468 & 0.0918 \\
30015-01-24-00 & 92.7 & Aug 18 16:22 & 16:54 & 01234 & C--A & 0.7396 & 0.0880 \\
30015-01-25-00 & 95.5 & Aug 21 11:33 & 12:13 & 01234 & C--A & 0.7405 & 0.0883 \\
\enddata
\tablenotetext{a}{Observation ID number assigned by the \xte\ project for the
	``C'' pointings.}
\tablenotetext{b}{Time Since 1998 May 18 0 UT}
\tablenotetext{c}{All observations listed were carried out in 1998.}
\tablenotetext{d}{All 5 PCUs were operational during a part of this
	observation.}
\label{obslogxte}
\end{deluxetable}

\begin{deluxetable}{lr}
\tablecaption{Timings of Optical Spin Minima}
\tablehead{ \colhead{Observing Season} & \colhead{Spin Minima (HJD)} }
\startdata
1993	& 2449197.743 \\
1994	& 2449506.838 \\
1996	& 2450280.702 \\
1996	& 2450340.602 \\
1998	& 2450982.948 \\
1998	& 2451040.737 \\
1999	& 2451317.877 \\
2000	& 2451813.672 \\
2002	& 2452481.936 \\
\enddata
\label{spinmin}
\end{deluxetable}

\begin{deluxetable}{rrrllll}
\tablecaption{The \xte\ Campaign Results}
\tablehead{ \colhead{ObsID} & \colhead{Cycle No\tablenotemark{a}} &
          \colhead{HJD$_{mid-ecl}$} & \colhead{Ingress} & \colhead{Egress} &
          \colhead{Duration} & \colhead{Post-ecl. Flux\tablenotemark{b}} \\
           & & \colhead{(predicted)} & \multicolumn{2}{c}{(cycle)} &
	    \colhead{(s)} & \colhead{(ct/s/PCU)} }
\startdata
30015-01-01-00 & 12496 & 2450952.0664 & $-$0.0290  & +0.0278  & 688  & 3.68 \\
30015-01-02-00 & 12525 & 2450956.1332 & $-$0.0296  & +0.0271  & 687  & 4.66 \\
30015-01-03-00 & 12553 & 2450960.0598 &            &          &      & 2.33 \\
30015-01-04-00 & 12583 & 2450964.2669 & $-$0.0275: & +0.0288: & 682: & 1.32 \\
30015-01-05-00 & 12612 & 2450968.3337 &            & +0.0317: &      & 1.05 \\
30015-01-06-00 & 12641 & 2450972.4005 &            &          &      & 0.62 \\
30015-01-07-00 & 12668 & 2450976.1868 & $-$0.0269  &          &      & 0.63 \\
30015-01-08-00 & 12699 & 2450980.5341 & $-$0.0269  & +0.0285: & 671: & 0.84 \\
30015-01-09-00 & 12726 & 2450984.3204 & $-$0.0275  & +0.0297  & 693  & 1.91 \\
30015-01-10-00 & 12774 & 2450991.0517 & $-$0.0304: & +0.0291  & 721: & 3.12 \\
30015-01-11-00 & 12804 & 2450995.2587 & $-$0.0305  & +0.0291  & 722  & 2.13 \\
30015-01-12-00 & 12825 & 2450998.2037 & $-$0.0304: & +0.0284  & 713: & 1.86 \\
30015-01-13-00 & 12854 & 2451002.2705 & $-$0.0297  & +0.0280  & 699  & 2.56 \\
30015-01-14-00 & 12881 & 2451006.0568 & $-$0.0282: & +0.0276  & 676: & 4.22 \\
30015-01-15-00 & 12911 & 2451010.2639 & $-$0.0297  &          &      & 2.80 \\
30015-01-16-00 & 12930 & 2451012.9283 & $-$0.0305: & +0.0329: & 768: & 2.29 \\
30015-01-17-00 & 12961 & 2451017.2756 & $-$0.0265: & +0.0318  & 706: & 1.24 \\
30015-01-18-00 & 12979 & 2451019.7998 & $-$0.0314: & +0.0330: & 780: & 0.91 \\
30015-01-19-00 & 13009 & 2451024.0069 & $-$0.0304  & +0.0295: & 726: & 0.75 \\
30015-01-20-00 & 13038 & 2451028.0737 & $-$0.0264  &          &      & 0.36 \\
30015-01-21-00 & 13067 & 2451032.1405 & $-$0.0267  & +0.0319: & 710: & 0.84 \\
30015-01-22-00 & 13095 & 2451036.0671 & $-$0.0273  & +0.0292  & 685  & 1.65 \\
30015-01-23-00 & 13125 & 2451040.2741 & $-$0.0279  & +0.0301  & 703  & 4.36 \\
30015-01-24-00 & 13153 & 2451044.2007 & $-$0.0295: & +0.0300: & 721: & 3.11 \\
30015-01-25-00 & 13173 & 2451047.0054 &            &          &      & 1.74 \\
\enddata
\tablenotetext{a}{Cycle number of the eclipse observed, as defined by
	our ephemeris (1).}
\tablenotetext{b}{Average count rate during orbital phase 0.04--0.09.}
\label{xteecl}
\end{deluxetable}

\end{document}